\documentstyle[12pt,aasms4]{article}
\newcommand{\pks}{PKS\,2155$-$304}
\newcommand{\bls}{BL Lac objects}
\newcommand{\bl}{BL Lac object}
\newcommand{\sax}{{\it Beppo}SAX}

\newcommand{\lcs}{light curves}
\newcommand{\lc}{light curve}
\newcommand{\etal}{et al.\,}

\newcommand{\cca}{cross-correlation analysis}
\begin{document}


\lefthead{Y.H.\ Zhang et al.}
\righthead{Rapid X-ray Variability of \pks\ }

\title{Rapid X-ray Variability of the BL Lacertae Object \pks\ }

\authoraddr {You Hong Zhang, SISSA/ISAS, Via Beirut 2-4, I-34014 Trieste, 
             Italy, e-mail: yhzhang@sissa.it}
\author {Y.H. Zhang \altaffilmark{1},
A. Celotti \altaffilmark{1},
A. Treves \altaffilmark{2},
L. Chiappetti \altaffilmark{3},
G. Ghisellini \altaffilmark{4},
L. Maraschi \altaffilmark{5}, \\ 
E. Pian \altaffilmark{6},
G. Tagliaferri \altaffilmark{4}, 
F. Tavecchio \altaffilmark{5},
C.M. Urry \altaffilmark{7} 
}
\altaffiltext{1}{International School for Advanced Studies, SISSA/ISAS,
  via Beirut 2-4, I-34014 Trieste, Italy; yhzhang@sissa.it,
  celotti@sissa.it}
\altaffiltext{2}{Dipartimento di Scienze, Universit\`a dell'Insubria,
Polo di Como, via Lucini 3, I-22100 Como, Italy; treves@uni.mi.astro.it}
\altaffiltext{3}{Istituto di Fisica Cosmica G.Occhialini, IFCTR/CNR, 
  via Bassini 15, I-20133 Milano, Italy; lucio@ifctr.mi.cnr.it}
\altaffiltext{4}{Osservatorio Astronomico di Brera, via Bianchi
  46, I-22055 Merate, Italy; ghisellini@merate.mi.astro.it;
  tagliaferri@merate.mi.astro.it}
\altaffiltext{5}{Osservatorio Astronomico di Brera, via Brera 28,
   I-20121 Milano, Italy; maraschi@brera.mi.astro.it;
   fabrizio@brera.mi.astro.it}
\altaffiltext{6}{Istituto TESRE/CNR, via Gobetti 101, I-40129 Bologna,
   Italy; pian@tesre.bo.cnr.it}
\altaffiltext{7}{Space Telescope Science Institute, 3700 San
   Martin Drive, Baltimore, MD 21218, USA; cmu@stsci.edu}


\begin{abstract}
We present a detailed power density spectrum and cross-correlation
analysis of the X-ray light curves of the \bl\ \pks\, observed with
\sax\ in 1997 (SAX97) and 1996 (SAX96), aimed at exploring the rapid
variability properties and the inter-band cross correlations in the
X-rays. We also perform the same analysis on the (archival) X-ray \lc\
obtained with ASCA in 1994 (ASCA94).

No large amplitude variability event on timescale of less than $\sim$
1 hour is found, and \lcs\ in different energy bands are highly
correlated with a tendency for the amplitude of variability to
increase with energy. The amplitude of variability is larger and the
(fastest) timescale shorter as the source becomes
brighter. Furthermore, while the average power density spectra of all
the \lcs\ present pronounced, featureless, {\it red noise } spectra,
with the power decreasing towards high temporal frequencies, the power
law slopes are somewhat different, indicating different variability
properties among the three observations.

We perform a cross correlation analysis using Monte Carlo simulations
in order to estimate the uncertainties on time lags between different
bands. A significant soft lag of $\sim 4$ hours between the 0.1$-$1.5
and 3.5$-$10 keV bands is seen in SAX96. During SAX97, in which the
source showed a high X-ray state correlated with a highly active phase
in gamma-rays, a short soft lag ($\sim 0.4$ hours) is detected between
the same energy bands. In contrast, ASCA94 presents an intermediate
soft lag of about 0.8 hours. These findings indicate that the
inter-band soft time lags are variable and suggest that the lag is
longer when the source is fainter. The time dependence of variability in
\pks\ is briefly discussed within a homogeneous synchrotron scenario for
blazars.

\end{abstract}

\keywords{BL Lacertae objects: individual (\pks) --- 
          galaxies: active ---
	  X-rays: galaxies 
	 }
\setcounter{footnote}{0}
\section{Introduction}

BL Lacertae objects represent a subclass of Active Galactic Nuclei (AGNs), 
emitting non-thermal radiation from radio to gamma-rays, even up to TeV 
energies. A defining property of \bls\ is that 
the radiation is strongly variable from radio to gamma-rays on different 
timescales. The mechanisms responsible for the non-thermal emission over 
such a wide energy range are commonly believed to be synchrotron and 
inverse Compton scattering from plasma in a relativistic jet oriented at 
a small angle with respect to the line of sight.

\pks\ is the brightest \bl\ at UV wavelengths and one of the  
brightest in the X-ray band. It was detected in gamma-rays by the EGRET  
experiment on CGRO (Vestrand, Stacy \& Sreekumar 1995; Sreekumar \&
Vestrand 1997), and it is one of the few BL Lacs observed at TeV energies 
(Chadwick \etal 1999). Its broad band spectrum shows two peaks: 
the first one is synchrotron emission peaking at UV and/or soft X-rays as  
most X-ray selected \bls\ (High frequency peak \bls\ , HBLs; Padovani 
\& Giommi 1995). The other one is around the gamma-ray region and it is 
attributed to Compton scattering by the same high energy electrons which 
are radiating via synchrotron. It has a very hard gamma-ray spectrum in
the 0.1$-$10 GeV region, with a power-law spectral index of
$\alpha_{\gamma} \sim 0.71$ (Vestrand \etal 1995), and a time-averaged  
integral flux of $4.2 \times 10^{-11}$ erg ${\rm cm^{-2} s^{-1}}$ above  
300 GeV (Chadwick \etal 1999). 

\pks\ has been one of the best targets of multiwavelength campaigns
because of its brightness. This kind of study has proved to be a
powerful tool to constrain radiation models through the study of
correlated variability among different bands. The first
multiwavelength campaign was performed, from radio to X-ray
wavelengths, in 1991 November, by ROSAT, IUE and ground-based
telescopes, and correlated variability was observed between UV and
soft X-rays with the UV lagging by $\sim 2$ hours (Edelson \etal
1995).  However, the source showed a definitely different variability
behavior in the 1994 May campaign based on IUE, EUVE and ASCA data.
Correlated variability was observed with larger amplitude at shorter
wavelengths, and significant soft lags, i.e. the UV lagging the EUV by
1 day, and the EUV in turn lagging the X-rays by 1 day (Urry \etal
1997).

Variability can be characterized by the power density spectrum (PDS
hereafter) and inter-band correlations: the PDS slopes and the
measured time lags impose strong constraints on radiation models. The
three long duration and high time resolution observations by \sax\ and
ASCA are rather suitable to carry on temporal studies. In the present
paper we perform detailed timing analysis for these observations, and
compare the results. Preliminary cross correlation analysis with Monte
Carlo simulations were firstly reported in Treves et al. (1999).

We briefly summarize the observations in section 2; the \lcs\ and
variability analysis are presented in section 3, followed by the PDS
analysis in section 4; in section 5 we carry out comprehensive \cca\
with detailed Monte Carlo simulations to determine the uncertainties
on inter-band lags.  The physical implications of the results are
discussed in section 6 and conclusions are drawn in the final section
7.

\section{Observations}

The \sax\ payload (Boella \etal 1997a) consists of four Narrow Field
Instruments (NFIs) which point in the same direction, namely one Low
Energy Concentrator Spectrometer (LECS) sensitive in the 0.1$-$10 keV
range (Parmar \etal 1997), and three identical Medium Energy
Concentrator Spectrometers (MECS) sensitive in 1.5$-$10 keV band
(Boella \etal 1997b).  Both the LECS and MECS detectors are Gas
Scintillation Proportional Counters (GSPC) and are in the focus of the
four identical X-ray telescopes. There are two more collimated
instruments: the High Pressure Gas Scintillation Proportional Counter
(HPGSPC) (Manzo et al. 1997) and the Phoswich Detector System
(Frontera et al. 1997), which are not however suitable to perform
temporal analysis because of the high background and limited
statistics on a source like \pks.  Therefore, for the following
timing analysis, only LECS and MECS data are used.

\sax\ NFIs observed \pks\ for more than 2 days during the Performance
Verification phase on 20-22 November 1996 (SAX96), and for slightly
less than 1.5 days during our AO1 observation from 22 to 24 November
1997 (SAX97). The effective exposure times for MECS and LECS were 63
ks and 22 ks for SAX97, and 108 ks and 36 ks for SAX96,
respectively. \sax\ data reduction procedure is described in detail by
Chiappetti \etal (1999). The \lcs\ were firstly presented by Giommi
\etal (1998) and Chiappetti \etal (1999). In particular in the latter
work the presence of a soft lag of about 10$^3$ s and a tendency of
the amplitude of variability to increase with energy have been found
in the SAX97 data.  

\pks\ was also monitored by the ASCA satellite for more than two days
from 19 to 21 May 1994 (ASCA94) coordinated with a multiwavelength
monitoring from radio to X-rays (Pesce \etal 1997; Pian \etal 1997;
Urry \etal 1997). ASCA includes two SIS and two GIS focal-plane
detectors (Tanaka, Inoue, \& Holt 1994). The X-ray light curve
considered here -- retrieved from the archive -- was taken from the
GIS detectors. Preliminary results were presented by Makino \etal
(1996).

In this paper we will perform detailed temporal analysis of the
different observations and compare the relative results. The log of
the three observations is shown in Table~\ref{tab:log}.

\section{Variability Analysis}

We analyze the light curves with the timing analysis software package  
XRONOS (Stella \& Angelini 1993). Unless otherwise specified, we {\it a
priori} separate the energy ranges into the following three bands: (1)  
0.1$-$1.5 keV as soft energy band referred to as LE band; (2) 1.5$-$3.5  
keV as the first medium X-ray band which we refer to as ME1 band; (3)  
3.5$-$10 keV as the second medium energy band, i.e. ME2 band. Note that
the LE band of the ASCA observation is 0.5$-$1.5 keV.

The light curves binned over 1000 or 2000s are shown in Figures 1,2
and 3 for the SAX97, SAX96 and ASCA94, respectively. We compute the
hardness ratios (HR) of ME1 to LE (HR1) and of ME2 to ME1
(HR2), which are also presented in the same figures.

\subsection{Variability parameters}

In order to quantify the variability properties, here we summarize the
general definition of the fractional $rms$ variability parameter
$F_{var}$ (e.g. Rodr\'{\i}guez-Pascual \etal 1997). The data series
$F_{i}(t)$ of the \lc\ has a standard deviation $\sigma^{2}_{F} =
\frac{1}{N - 1 } \sum_{i=1}^{N} (F_{i}(t) - \overline{F})^{2}$, where
$\overline{F}$ is the mean count rate. In addition, we define the {\it
expected variance}, due to random errors $\sigma_{i}(t)$ associated
with $F_{i}(t)$, as $\Delta^{2}_{F} = \frac{1}{N} \sum_{i=1}^{N}
\sigma^{2}_{i}(t)$.  The {\it excess variance}, $\sigma_{exc}$, is
then defined as the difference between the standard deviation
$\sigma^{2}_{F}$ and the expected variance $\Delta^{2}_{F}$, i.e.
$\sigma^{2}_{exc} = \sigma^{2}_{ F} - \Delta^{2}_{F}$, from which we
can define the fractional $rms$ variability parameter as $F_{var} =
\sigma_{exc}/\overline{F}$.

The above parameters only characterize the mean variability of a
source.  However, a direct measurement of the fastest timescale on
which the intensity can change is crucial as it may constrain the
source size, and thus luminosity density, accretion efficiency or
beaming parameters, and black hole mass. This requires to identify
rapid variability events rather than the average variability
properties. One often considers the so-called ``doubling time'' as a
reasonable measure of the fastest and meaningful timescale of a source
(e.g. Edelson 1992). More precisely, here we define the ``doubling
time'' as $T_{2} =|\frac{F\Delta T}{\Delta F}|$, where $\Delta T =
T_{j} - T_{i}$, $\Delta F = F_{j} - F_{i}$, and $F = (F_{j} +
F_{i})/2$, and consider the minimum value of $T_{2}^{ij}$ over any
data pairs as the shortest timescale for each observation, keeping in
mind that this quantity is ill-defined, strongly depending on sampling
rate, length and signal-to-noise ratio of the observation (Press
1978).  The error on $T_{2}^{ij}$ is propagated through the errors on
the fluxes $F_{i}$ and $F_{j}$, and {\it a priori} we neglect the
value of $T_{2}^{ij}$ if the error is larger than $20\%$. 
  
The variability parameters defined above are reported in
Table~\ref{tab:var}.

\subsection{Results}

\subsubsection{SAX97}

Figure~\ref{toolc} presents the light curves and hardness ratios.  At
the beginning of the observation \pks\ exhibited a large flare, with a
variation by a factor $\sim 4$, followed by two other events of
smaller amplitude.  The second flare presents similar rising and
declining timescales. As shown in Table~\ref{tab:var}, the variability
amplitude is to some extent different in the three bands, increasing
with increasing energy ($F_{var}$ is 0.22, 0.27 and 0.30 in the LE,
ME1 and ME2 bands, respectively). No variations on timescales of less
than $\sim$ 1 hour are found. The most rapid variation event -- the
fastest among the three observations -- occurred during the first
flare, with minimum values of $T_{2}$ of about 3.4, 1.9 and 1.8 hours
in LE, ME1 and ME2 bands, respectively. We notice that these
timescales are much shorter and the fluxes about $50\%$ higher than
those of SAX96 (see Table~\ref{tab:var}), indicating faster variability
with higher intensity.

From the last two panels of Figure~\ref{toolc}, one can see that the
HR1 presents a global trend similar to that of the intensities (see
Chiappetti \etal 1999 for more details). However, no
statistically significant correlation seems to be present, as HR1 has
the same value during the first two peaks which have significantly
different intensities and is smaller during the end of the
observation, although the average intensity is similar to that of the
second peak.  HR2 does not show any trend.

\subsubsection{SAX96}

As shown in Figure~\ref{pvlc}, an approximately symmetric flare was
seen in the middle of the observation, which is well resolved with
similar rising and decaying time scales. A flare of lower intensity is
visible at the beginning, while a larger flare probably occurred
towards the end of the observation, although the observation is
incomplete.  Some small-amplitude variability is also detected. 

The $F_{var}$ are comparable for the ME1 and ME2 bands ($\sim 0.13$),
and are $\sim 30\%$ larger than that relative to the LE band. The
estimated ``doubling times'' are about 22, 14 and 8 hours for the LE, ME1
and ME2 bands, respectively. From Figures 1,2 and Table~\ref{tab:var}, it
is clear that during SAX96 \pks\ was in a relatively faint state with
smaller amplitude and longer timescale variability, compared to SAX97.

The hardness ratio HR1 shows a behavior similar to that of the light
curves, in the sense that the spectrum is harder at higher
intensities, while again HR2 does not follow any trend (see
Figure~\ref{pvlc}).

\subsubsection{ASCA94}

The light curves and hardness ratios relative to this observation are
plotted in Figure~\ref{ascalc}. A large amplitude flare, with an
approximately symmetric shape, is clearly seen at the beginning of the
observation although the rising portion of the event is not fully
sampled.  \pks\ was more variable in this period than during the other
two observations, as can be seen from $F_{var}$ (Table~\ref{tab:var}),
with a flux intermediate between SAX97 and SAX96. The estimated
``doubling times'' are about 5.6, 4.8, and 4.5 hours in the LE, ME1
and ME2 bands, respectively.

A significant characteristic of ASCA94 is that the hardness ratios
present a trend of linear decrease over the whole period, which is a
general signature that the spectra become softer when the source is
fainter. Urry \etal (1997) showed the same trend through the spectral
fits.

\section{PDS Analysis}

AGN variability can be statistically characterized by its PDS.  The
PDS of very few Seyfert galaxies and \pks\ generally behave as power
laws, proportional to $f^{- \alpha}$ over some temporal frequency
range, where $f$ is the temporal frequency (e.g. Edelson \& Nandra
1999; Hayashida \etal 1998; Tagliaferri \etal 1991). For \pks , the
durations of the observations considered here are much longer ($\sim2$
days) than previous ones (e.g. EXOSAT), allowing us to determine the
PDS over a range extending towards relatively lower frequencies.
Because of low exposure efficiency of the LECS ($\sim 20\%$), here we
focus on the \sax\ MECS and ASCA \lcs\ in the 1.5$-$10 keV region.

The PDS analysis is carried out with the direct Fourier transform
algorithm which is included in the timing series analysis package
XRONOS. For these observations, the PDS is calculated for the
background-subtracted light curves with 10s time resolution, as each
PDS in our cases approaches (white) noise level before $\sim 10^{-2}$
Hz, clearly smaller than the Nyquist frequency of $5 \times 10^{-2}$
Hz at 10s bin size. The average count rate is subtracted from the bins
before the PDS is calculated. In order to improve the signal-to-noise
and study the mean variability properties of \pks , the light curves
are divided into several short intervals with each interval sampling
4096 points. The SAX97 \lc\ presents 3 good intervals; the SAX96 \lc\ has 4
good intervals, while we neglect the last part of the \lc\ which contains
a long interruption towards the end of the observation; the ASCA94
observation is divided into 4 good intervals. For each \lc , the
power spectra from each interval are then averaged. 

An important issue, discussed in detail by Tagliaferri \etal (1991),
is the data gap filling, which is unavoidable for a low orbit X-ray
satellite.  The gap-filling procedure could strongly affect the
derived PDS slope, artificially increasing the power at high
frequencies and introducing spurious quasi-periodic oscillations
(QPOs) (Tagliaferri \etal 1996). In order to decrease the effect of
data gaps in determining the PDS, we adopt the gap filling procedure
defined as ``running mean gap filling'' in XRONOS. This method
replaces the data gaps with the moving average of the light curve
calculated in our cases over a duration of about 1.5 hour. In this
way, the gaps are bridged in a smooth way, which not only simulates
real events but also reduces the bias introduced by the window
function. We also determined {\it a posteriori}, by considering a
somewhat different duration (e.g. 2 hours), that the slope of the PDS
is rather insensitive to the filling duration over which the running
mean is calculated. This indicates that the running mean could follow
the light curve behavior on time scales of hours as long as the gap
filling duration is substantially shorter than the whole interval and
longer than the data gaps, so that the gap center could be ``linked''
with the running mean from a relatively high number of points and the
statistical fluctuations are reduced.

The average PDS (after average noise subtraction) obtained in this way
from each individual observation are shown in Figure~\ref{fig:pds}a,b,c
for SAX97, 96 and ASCA94 \lcs , respectively. These are rebinned in
logarithmic intervals of 0.18 (factor 1.5) to reduce the noise and
allow the estimation of error bars. This means that the first point is
still the lowest frequency point, but the second point is derived by
averaging the next two points, etc. In such a way the PDS appear
nearly equi-spaced in a log-log diagram. Each PDS is normalized so
that its integral gives the squared $rms$ fractional variability
$F_{var}$ (therefore the power spectrum is in unit of $\rm {F_{\it
var}^{2}/Hz}$), which is normalized to the squared average count rate.
The expected (white) noise power level must be subtracted to obtain
the $F_{var}$ of the light curve (this level is about 1.5, 1.6 and 1.2
for SAX97, 96 and ASCA94 data, respectively).  The error bars
represent the standard deviation of the average power in each rebinned
frequency interval, where the power in each bin is $\chi^{2}$
distributed with $2N$ degrees of freedom, where $N=ML$ is the total
number of points used to produce the mean power in each frequency bin
(from $M$ intervals and $L$ independent Fourier frequencies).

From Figure~\ref{fig:pds}, one can see
that each PDS shows a strong {\it red noise} spectral component which
decreases with increasing frequencies, without any significant narrow
feature that would be indicative of periodic or quasi-periodic
variability. This component approaches the noise level at $\sim 6
\times 10^{-3}$ Hz for the SAX97, 96 \lcs , and at $\sim 1 \times
10^{-3}$ Hz for the ASCA94 data set. In addition, we note some
differences among the three PDS.  The SAX97 PDS clearly shows more
power than the SAX96 one at lower frequencies, indicating a
flatter PDS for SAX96 (this is easy to reconcile with the fact that
$F_{var}$ of SAX96 ($\sim 0.13$) is much less than that of SAX97
($F_{var} \sim 0.3$)).  The ASCA94 PDS has much less power than that
of SAX97 over the whole range of temporal frequencies considered here,
consistent with the fact that the ASCA94 PDS approaches the noise
level at relatively lower frequencies. However, this does not agree
with their corresponding $F_{var}$ values. Let us consider the origin
of this discrepancy. The SAX97, 96 \lcs\ show more or less identical
amplitude of variability over the whole observations, i.e.  similar
$F_{var}$ for each interval, while the ASCA94 \lc\ does not present
pronounced variability after the large flare at the beginning of the
observation. Thus $F_{var}$ for the ASCA94 dataset significantly
changes from one interval to another, being about 0.23 and 0.11 for
the flare and (almost) constant flux intervals of the ASCA94 \lc ,
respectively, but 0.35 when calculated over the whole duration. This
makes the ASCA94 data set more variable if we consider it as a
whole.  For this reason we firstly compute the PDS in different
intervals and normalized it so that the integral gives its own squared
$F_{var}$ value, and then obtained the average PDS by averaging the
power spectra from each interval. Because the \lc\ of ASCA94 is
characterized by different $F_{var}$, we should use a mean value
averaged from the (four) intervals considered in deriving the average
PDS. This average $F_{var}$, which is much smaller than those of the
SAX97, 96 \lcs , indeed well agrees with the average PDS. Note that
instead, for the SAX97 and SAX96 \lcs\ , the average $F_{var}$ from each
interval is identical to that of the whole observation.
     
In order to quantify the slope of the PDS, power law model is fitted
to each average power spectrum in the frequency interval $\sim 6 \times
10^{-5}$ to $\sim 6 \times 10^{-3}$ (SAX97, SAX96) or to $\sim 1.5 \times
10^{-3}$ Hz (ASCA94). The lowest frequency point of each PDS was ignored
because they tend to be more noisy, and also for comparison with previous
PDS analysis.  The best-fit power law slopes are $\sim 2.2, 1.5$ and 2.2
for SAX97, 96 and ASCA94 PDS, respectively. We also compute in the same
way the average PDS after the removal of a linear trend from the \lcs ,
in which the power law slopes for the ``de--trended'' PDS are consistent
with the above values within $1\sigma$, respectively.  The fitting details
are shown in Table~\ref{tab:pds}. 

\section{Cross Correlation Analysis}

The first clear result visible from the \lcs\ is that the variations
in the different X-ray bands are all correlated. Indeed, these
intensive monitorings with high time resolution and long duration
allow detailed measurements of the inter-band cross correlation
properties, and in particular to make quantitative estimates of the
degree of correlation and of any lags between variations at different
X-ray wavelengths. Two cross correlation methods, namely the Discrete
Correlation Function (DCF) and Modified Mean Deviation (MMD), are
applied. In the following, a positive lag indicates the higher energy
X-rays leading the lower energy ones, while negative indicates the
opposite.

\subsection{Cross Correlation Analysis Technique}

\subsubsection{DCF}

The DCF is analogous to the classical correlation function (e.g. Press
et al. 1992) except it can work with unevenly sampled data.  The DCF
technique was described in detail by Edelson \& Krolik (1988) and
applied to \pks\ by Edelson \etal (1995) and Urry \etal (1997) to
measure the time lags between UV and X-ray during the two
multiwavelength campaigns mentioned above. Here we bin the original light
curves and fix the DCF resolution according to the following criteria:
(1) the bin sizes in both the \lcs\ and the DCF should be at least 3
times smaller than any possible lag;
(2)the bin size should also be as large as possible to reduce 
the error on the DCF.
The resulting DCFs, only for the LE/ME2 case, are shown in
Figure~\ref{fig:ccf}a,c,e
(left panels) for the three observations, respectively. In order to
quantify any time lag, we fit
the DCF with a Gaussian function plus a constant, and take the
Gaussian centroid, rather than the DCF peak, as the lag between the
two energy bands (see arguments by Edelson \etal 1995 and Peterson
\etal 1998). The two main advantages of this are:
(1) the Gaussian fit takes into account the overall symmetry of the
cross-correlation distribution around the peak, reducing the
possibility of spurious lags due to a particular DCF point that could
originate from statistical errors; (2) we found that -- under the two
conditions mentioned above -- the lag and its uncertainty derived from
a Gaussian fit are insensitive to the bin sizes of both the \lcs\ and
the DCF.
The Gaussian fits to the DCFs are also shown in the same figures, and
Table~\ref{tab:lag} reports the fitting results for the LE/ME1, ME1/ME2
and LE/ME2 cases. It should be noted that a Gaussian fit -- although
representative of the peak position and dispersion for both the DCF and
MMD -- does not necessarily provide a statistically adequate fit to these
functions.

\subsubsection{MMD}
In order to check the results suggested by the DCF technique, we
perform a similar analysis by using the MMD method introduced by
Hufnagel \& Bregman (1992). The MMD considers the mean deviation of
the two cross correlated time series as the correlation estimator and
the minimum value of the MMD should correspond to the best correlation
point (lag).  Thus, unlike the DCF, it cannot be used to estimate the
significance of the correlation between different wavelengths.  As
with the DCF, we take the centroid of a Gaussian fit as the measured
lag. The MMD results with their Gaussian fits, only for the LE/ME2 case, 
are shown in Figure~\ref{fig:ccf}b,d,f (right hand panels), and
the results of the fits are reported in Table~\ref{tab:lag} for the
LE/ME1, ME1/ME2 and LE/ME2 cases.

\subsection{Monte Carlo Simulations}

As suggested by Peterson \etal (1998), the uncertainty on the
cross-correlation lag are dependent on both the flux uncertainties in
individual measurements and the observational sampling uncertainties
of the light curves. So, the statistical significance of the detection
of a lag can not be assessed just by a cross-correlation analysis. In
order to test the dependence of our findings on photon statistics, in
this section we apply to our data the model-independent Monte Carlo
simulation method introduced by Peterson \etal (1998). Because of the
uncertainties just mentioned, the method considers ``flux
randomization'' (FR) and ``random subset selection'' (RSS). FR assumes
that the errors on fluxes resulting from the total photon number in a
bin (several hundred photons in our cases) are normally
distributed. Thus FR just takes each real flux $F_{i}$ and modifies it
by adding a random Gaussian deviation based on the quoted error
$\sigma_{i}$ for each data point of the light curves. So, the
modification of each data point is statistically independent of each
of the others, and therefore the dependence of lags on flux errors can
be easily assessed through the FR simulations. RSS tests instead the
sensitivity of a cross correlation lag by considering only subsets of
the original light curves with no dependence on previous selection but
still preserving the temporal order of the data points. The
probability of random removal of any data point is $\sim 1/e \simeq
0.37$ which is a Poisson probability. Thus each RSS realization is
based on a randomly selected subset which is typically $\sim 37\%$
smaller than the real data set. Peterson \etal (1998) argue that RSS
gives a fairly conservative estimate of the uncertainties due to
sampling. We thus take the combination of FR and RSS in a single
simulation to test together the sensitivity of the cross-correlation
lags on flux uncertainties and sampling characteristics. We apply the
DCF and the MMD to each FR/RSS Monte Carlo realization to determine
individual lags obtained from the centroid of the Gaussian fit to each
independent realization. The same process is repeated 2000 times to
build up a cross-correlation peak distribution (CCPD; Maoz \& Netzer
1989), which is not necessarily a normal distribution (e.g. Peterson
\etal 1998). The CCPDs for the three observations are displayed in
Figures 6$-$8 (their different widths result from the different
photon statistics), respectively.  From the CCPD we can determine the
probability that a given lag falls in some particular likelihood range.
In our cases (2000 realizations), we determine the lower (upper) extrema
of the $68\%$ and $90\%$ confidence ranges by taking the 320th (1680th)
and 100th (1900th) smallest values from all realizations,
respectively.  The results of the simulations are shown in
Table~\ref{tab:mcs}. In addition, we tested that the results are
insensitive to the bin sizes of both the light curves and the
cross-correlation.

In Figure~\ref{bijiao} the lags derived from the DCF/MMD methods
directly and through the simulations are compared: indeed they are
fully consistent within the uncertainties estimated from the FR/RSS
simulations.

\subsection{Results}

\subsubsection{SAX97}

We remind that during SAX97 the source was in a relatively high state
compared to SAX96, and variability was more pronounced.  The
inter-band correlation coefficients $r_{0}$ (see Table~\ref{tab:lag})
indicate that the X-rays in different bands are highly correlated. The
cross correlation analysis show a very short soft lag between the LE
and ME2 bands ($\sim$ 1000 s), while the lags for LE/ME1 and ME1/ME2
are consistent with zero (see Figure~\ref{fig:ccf}(a,b) and
Table~\ref{tab:lag}).  The FR/RSS Monte Carlo simulations confirm
these findings with high significance (see Figure~\ref{tooccpd} and
Table~\ref{tab:mcs}).

\subsubsection{SAX96}

The values of $r_{0}$ derived for LE/ME1, ME1/ME2 and LE/ME2
correlations (see Table~\ref{tab:lag}) suggest that the \lcs\ are also
strongly correlated during this relatively faint state. However, in
contrast to SAX97, we find significant soft X-ray lags relative to
higher energy X-rays. It is apparent from Figures~\ref{fig:ccf}(c,d),
and ~\ref{pvccpd} and Tables 4,5 that the lags estimated with the DCF and
MMD methods are compatible within the uncertainties of the FR/RSS Monte
Carlo simulations, indicating the presence of a soft positive lag of $\sim
4$ hours between the LE and ME2 bands. Soft lags of about $2$ hours are
also shown by the LE/ME1 and ME1/ME2 cross correlation functions. Note
also that the soft X-ray lags in this case are the largest recorded so far
for \bls\ in the X-rays.

\subsubsection{ASCA94}

The state of the source during the ASCA94 observation is intermediate
among the two $Beppo$SAX observations.  Also these data show strong
correlations among the different bands (see Table~\ref{tab:lag}). The
DCF and MMD analysis (see Figure~\ref{fig:ccf}(e,f) and
Table~\ref{tab:lag}) reveal soft lags intermediate between those of
the SAX97 and SAX96 data sets: the LE lags the ME2 by about $0.8$
hours, while LE lags ME1 and ME1 lags ME2 by $\sim$ 0.4 hours.  These
results are also confirmed by the FR/RSS Monte Carlo simulations (see
Figure~\ref{ascaccpd} and Table~\ref{tab:mcs}).

\subsubsection{Comparisons}

The results from the three observations corresponding to different
intensity states of \pks\ suggest that the soft time lags are variable
and possibly related to the source intensity, the soft lag becoming
larger when the source is fainter. We illustrate this behavior in
Figure~\ref{correlation}a, where the lags between LE and ME2 are
plotted against the mean fluxes in the ME2 band.  A similar behavior
is also present between the lags and the ratios of the maximum to the
minimum count rate (see Figure~\ref{correlation}b).  As a comparison,
in Figure~\ref{correlation}c, we include the upper limit to the soft
lag between the 0.1$-$2 and the 3$-$6 keV bands obtained from the
EXOSAT observation of 24 Oct 1985 (Tagliaferri \etal 1991). This
figure shows a power law relation (logarithmic axis) between the lags
and the fluxes. This suggestive trend might give crucial clues on the
emission processes and physical parameters in \pks, and strongly
requires the comparison with the results of time-dependent emission
models.

We also notice that the lags ($\tau$) are qualitatively
anti-correlated with the correlation coefficients ($r_{0}$) between
the different energy bands. From Table~\ref{tab:lag}, it can be seen
that the $r_{0}$ of SAX97 and ASCA94 are significantly higher than
those of SAX96.  On the contrary, the soft lags of SAX97 and ASCA94
are smaller than those of SAX96. This anti-correlation is indeed
expected. If obvious soft lags were present, the maximum amplitude of
the cross correlation function would be significantly shifted away
from the zero lag point and thus the standard correlation coefficient
$r_{0}$ would obviously decrease if the cross correlation function
were a smooth Gaussian function (but not necessarily). Therefore,
variable lags in different states are qualitatively suggested by the
variable correlation degrees without any measured time shift. This
behavior occurs also between the $\tau$ and $r_{0}$ within the
different inter-band correlations (for example, the lags become larger
with $r_{0}$ becoming smaller in SAX96).
      
Interestingly, variations of the soft lags found here in the X-ray
bands are reminiscent of the variations of the UV lags with respect to
the X-ray (from $\sim 2$ hours to $\sim 2$ days) between the 1991 and
1994 intensity states of this same source (see Introduction).

\section{Discussion}

The high degree of correlation and time lags between variations at
different wavelengths provide strong constraints on the physical
parameters of blazars. The previous multiwavelength monitoring
campaigns of \pks\ found different variability behaviors (Edelson
\etal 1995; Urry \etal 1997).  In particular, the 1991 campaign showed
the soft X-ray leading the UV by just $\sim 2$ hours (the result of
the cross correlation analysis was recently confirmed by Peterson
\etal (1998) on the basis of simulations similar to those used in this
paper). However, during the 1994 campaign the UV lagged the X-ray by
$\sim 2$ days. Time lags of the soft compared to the hard X-rays were
suggested by the ASCA observations of \pks\ and MKN 421 (Makino \etal
1996; Takahashi \etal 1996), while EXOSAT observations of \pks\ showed
no evidence of lags, with upper limits of a few hundred seconds
(Tagliaferri \etal 1991).

In order to interpret the inter-band variable time lags, the
development of time dependent models taking into account the effects
of particle injection/acceleration, cooling and diffusion would be
required. However, the time dependent problem is in general very
complicated and only some simplified and specific cases have been
considered so far. Mastichiadis \& Kirk (1997) showed that,
within the assumptions of an homogeneous SSC model, an increase in the
maximum energy of the injected electron population can reproduce the
rapid X-ray flares as well as the spectral evolution of blazars like
MKN 421.  Interestingly, they also show that these features cannot be
due to changes of both the magnetic field and the amount of injected
electrons. In addition, Chiaberge \& Ghisellini (1999) pointed out the
importance, for both spectral evolution and time lags, of delays due
to light crossing the radiating region. This effect is superposed to
the wavelength dependent timescales due to the different cooling times
of radiating electrons. In contrast to the above studies,
Georganopoulos \& Marscher (1998) modelled, using a time-dependent
inhomogeneous accelerating jet model, the evolution of flares during
the two multiwavelength campaigns on \pks\ . Within this scenario, the
different variability features could be reproduced by assuming that
plasma disturbances with different physical properties occur in an
underlying jet characterized by the same physical parameters. The
small time lag between the UV and X-ray bands in the 1991 November
campaign would indicate quasi-co-spatiality of the regions radiating at
these frequencies, assuming an injected electron distribution similar
to that characterizing the underlying jet emission. However, the clear
time lag between these same bands in the 1994 May campaign are interpreted
as an indication of spatial separation of the emitting regions. The
separation can be due to the propagation downstream of the electrons
while progressively radiating at lower frequencies.  This however also
requires the injected electrons to be narrower in energy than during
the 1991 November event.

Clearly, in order to pin down the origin and nature of variations,
both systematic observational trends and a thorough analysis with the
different models are needed. In this work, we have concentrated on the
first aspect, but let us examine the simplest (and analytical)
considerations one can draw from the observational results. 

Obviously, the homogeneous synchrotron self--Compton model is the
simplest interpretation for the X-ray emission and overall spectral
energy distribution (SED) of \pks\ (e.g. Chiappetti \etal
1999). According to this picture a quasi-stationary population of
particles is responsible for a ``quiescent'' flux level, while flares
result from a uniform injection and/or acceleration of relativistic
electrons over a time interval $\Delta t$. The evolution of the
particle distribution is governed by the radiative cooling through
synchrotron emission which dominates in the X-ray band. As the
radiative losses are energy dependent, that is, radiative lifetime of
electrons is inversely proportional to the emitted frequency, low
energy photons are expected to lag high energy ones (e.g. Urry \etal
1997; Takahashi \etal 1996). In particular, within this simple
scenario, it is possible to relate the observed time lag to the
physical parameters of the source (see also Tavecchio, Maraschi \&
Ghisellini 1998), and in the observer's frame this can be expressed as

$$ 
B\delta^{1/3}=223.5 \left( \frac{1+z}{E_{l}} \right)^{1/3}  
\left[ \frac{1-(E_l/ E_h)^{1/2}}{\tau _{obs}} \right]^{2/3}  (G)
$$

\noindent
where $E_{l}$ and $E_h$ refer to the low and high X-ray energies (in
units of keV), and $\tau _{obs}$ (sec) is the observed lag between
$E_{l}$ and $E_h$ photons. Under the synchrotron cooling assumption,
the observed time lag $\tau _{obs}$ depends only on the magnetic field
intensity $B$ and the bulk Doppler factor $\delta$ of the radiating
region.  If we take $E_l$ as 0.8 keV, $E_{h}$ as 7 keV, and
$\tau_{obs}=4.0, 0.8$, and $0.4$ hours for each of the observations,
our results would imply $B\delta^{1/3}$ $\sim$ 0.32, 0.94 and 1.49 $G$
for SAX96, ASCA94 and SAX97, respectively. Interestingly, Chiappetti
et al. (1999) found that the model parameters derived through the
fitting of the broad band spectrum during the SAX97 observation are
consistent with those estimated from the observed soft lag.

A further piece of information is given by the trend between observed
soft time lags and fluxes (see Figure~\ref{correlation}). If we for
example assume that $\delta$ has not changed, the simplest scenario
would suggest that $B$ varied by a factor $\sim$ 5 from SAX96 to
SAX97. Although qualitatively consistent with the variation in the
flux, this can not quantitatively reproduce the observed
correlation. In fact, under the (simplistic) assumptions of variations
occurring only in the magnetic field, one would expect $F \sim
B^{1+\alpha _{x}}$ and $B\delta^{1/3} \sim \tau_{obs}^{-\frac{2}
{3}}$, thus implying that the relation between intensity and lag is
given by $F \sim \tau_{obs}^{- \frac{2(1+\alpha _{x} )}{3}}$ (assuming
$\delta$ constant), where $\alpha _{x}$ is the X-ray spectral index.
For $\alpha _{x} \sim 1.0$, we have $F \sim
\tau_{obs}^{-\frac{4}{3}}$.  For example, the change in the lag by a
factor $\sim 10$ from SAX97 to SAX96 would imply a flux variation of a
factor $\sim 22$. However, the corresponding observed flux just
changed by a factor $\sim 1.4$ in the 0.1$-$1.5 and 3.5$-$10 keV
bands.  The predicted relation between the lags and the fluxes under
this hypothesis is also shown in Figure~\ref{correlation}a,c, where it
can be clearly seen this is much flatter than that of the observed one
between all of the observations.  Therefore, as one might expect,
other physical quantities, such as the density and energy spectra of
the electron population and/or the Doppler factor, have to vary if the
observed relation between flux and lag holds within the homogeneous
synchrotron self--Compton scenario.

One more interesting piece of information for \pks\ is given by the
good correlation found between variability parameters and source
intensity. These have been plotted in Figure~\ref{correlation2}, as
the fractional variability parameter ($F_{var}$) and ``doubling time''
($T_{2}$) against the source flux in the 1.5$-$10 keV band: as the
source gets brighter, the average amplitude of variability is larger,
and the fastest variability timescale shorter. Although this is
only a suggestive trend - due to the limited statistics - it seems to
indicate that the properties characterizing variability are not
random: any mechanism(s) invoked to account for (variable) radiative
dissipation has to intrinsically produce this behavior.  More
observations with high time resolution are clearly required to confirm
and quantify this trend.

We also (qualitatively) stress that the flux--time lag
relationship could be associated with the importance of light crossing
effects with respect to the cooling timescales. A more intense flux
could be associated with relatively efficient dissipation,
e.g. occurring at a shock front, which for a quasi--planar geometry
(shock-in-jet model, in which a thin shock wave moves down a
cylindrically symmetric jet, Marscher \& Gear 1985) could imply that
light crossing effects do not dominate, and thus small time lags. A low
source state, more similar to the quiescent underlying jet emission
might be associated with an acceleration/injection of particles in a
larger region: due to significant light crossing effects, the observed
variability would be smoother and result in larger time lag between
different frequencies.

Finally, let us consider the information given by the PDS, which
statistically characterizes the variability of a source. It
should be noted that with the available data it is not easy to
determine through the PDS the typical (or minimum) variability
timescale, as the \lcs\ on small time bins becomes very noisy.  More
powerful techniques, e.g. the structure function, have to be
considered. In general, the amplitude of variability decreases as the
timescales become shorter. Previous studies, mainly of Seyfert
galaxies, show that their PDS can be approximated by power laws with
slopes ranging between 1.5 and 2.0 (e.g. McHardy 1999), providing
valuable constraints to discriminate among possible models. However,
the PDS of \bls\ has not yet been well studied in the X-ray band. The
best determined PDS has been derived by Tagliaferri \etal (1991) for
\pks\ using EXOSAT observations. An average power law slope of $\alpha
\sim 2.5$ was obtained for the PDS in the 1$-$6 keV band, which
however reduced to 1.9 after the removal of the linear trend (as
required in that case). Our analysis shows, over the same temporal
frequencies, that the slope of the average PDS from each observation
is consistent within $1\sigma$ with that of the ``de--trended'' PDS
derived from EXOSAT data. The fastest variability time scale inferred
from the PDS may reach $\sim 1000 s$, although this is largely
uncertain because of the noisy PDS. We note that this time scale
is consistent with that estimated from the PDS of the EXOSAT
observations (Tagliaferri \etal 1991).  Moreover, Paltani (1999)
recently determined a similar minimum time scale ($\sim 600s$) from
the EXOSAT data by using the structure function.  Interestingly, the
most rapid variability estimated from the ``doubling time'' in the
three observations occurred on a similar timescale ($\sim 1$ hour), at
least in SAX97. Of course, longer and uninterrupted X-ray monitoring
will be crucial for constraining the PDS of blazars.

\section{Conclusions}

We have considered three long-duration X-ray \lcs\ of \pks\ with high
time resolution, and performed detailed time series analysis on
them. The intensities in soft and medium X-rays are always well
correlated, but with significantly different soft lags, suggesting
that variability properties are time dependent and/or different
mechanisms responsible for the variability may be at work. The
three \lcs\ presented here, which are sampled over short timescales, do
not seem to show any direct correspondence with the overall/long term
variability properties of the source, as suggested by the two closely
similar ROSAT light curves with about 5 years separation 
discussed by Brinkmann \& Siebert (1999).

The most important conclusions presented in this paper can be
summarized as follows:

(1) \pks\ shows several well-defined symmetric X-ray flares with
similar rising and declining timescales. The amplitude of variability
increases with increasing frequency. Very rapid variability events are
not found on timescales of less than 1 hour;

(2) the average PDS of SAX97 has significantly more power than those
relative to the other two observations, indicating that \pks\ was more
variable in this period. In addition, the rapid timescales and average
amplitudes of variability may correlate with the source intensity, in
the sense that higher brightnesses correspond to shorter timescales
and larger amplitudes;

(3) the inter-band X-rays are highly correlated in all cases, but show
different soft time lags, possibly correlated with the source
intensity. During SAX96 the source was in a relatively low state and
showed the longest soft lag ($\sim 4$ hours) recorded so far at X-ray
wavelengths in \bls . The SAX97 \lcs , which correspond to a high
state, do not show significant time lag, while the ASCA94 \lcs\
present intermediate time lag;

(4) within the simple homogeneous synchrotron (self--Compton) model
for \pks , the time lags could be interpreted as related to the
cooling time scale of the relativistic emitting electrons, although
the simplest change in the field intensity cannot quantitatively
account for the observed dependence of lag on intensity;

(5) the variability of the (X--ray) inter-band soft time lags of \pks\
is reminiscent of variations of lags between the UV and X-ray bands
observed during the 1991 and 1994 multiwavelength campaigns (Edelson
\etal 1995; Urry \etal 1997).

\acknowledgments

The anonymous referee is thanked for constructive comments. YHZ and
AC acknowledge the Italian MURST for financial support. This work was
partly done in the research network ``Accretion onto black holes, compact
stars and protostars'' funded by the European Commission under contract  
number ERBFMRX-CT98-0195.


\clearpage

\figcaption[toolc.ps] {Light curves and hardness ratios of \sax\ 1997
November 22-24 observation.  Data are rebinned in 1000s. From top to
bottom panel: light curve in the 0.1$-$1.5, 1.5$-$3.5 and 3.5$-$10 keV
bands, respectively, and hardness ratio between the 1.5$-$3.5 and
0.1$-$1.5 keV bands (HR1) and between 3.5$-$10 and 1.5$-$3.5 keV bands
(HR2). Note that the temporal coverage of the LECS is much more sparse
than that of the MECS.  The simultaneous data points between LECS and
MECS are indicated by filled symbols. The HR1 shows a clear
correlation with the source brightness, not present for HR2.
\label{toolc} }

\figcaption[pvlc.ps]
{ Light curves and hardness ratios of \sax\ 1996 November 20-22 observation.
Data are rebinned in 2000s. The symbols have the same meaning of those in 
Figure 1. The HR1 (but not HR2) well correlates with the source intensity.
\label{pvlc} }

\figcaption[ascalc.ps]
{ Light curves and hardness ratios of ASCA 1994 May 19-21 observation. Data 
are rebinned in 2000s. The symbols have the same meaning of those of
Figure 1. Both the HR1 and HR2 closely follow the source intensity trend.
\label{ascalc} }

\figcaption[pds_a.ps, pds_b.ps, pds_c.ps]
{ PDS with the best power law fit (after average noise subtraction) in
the 1.5$-$10 keV band. (a) SAX97; (b) SAX96; (c) ASCA94. 
\label{fig:pds} }

\figcaption[ccf_a.ps,ccf_b.ps,ccf_c.ps,ccf_d.ps,ccf_e.ps,ccf_e.ps]
{ DCF and MMD between 0.1$-$1.5 and 3.5$-$10 keV. (a,b) SAX97; (c,d)
SAX96; (e,f) ASCA94. The best fit consists of a Gaussian
function plus a constant. 
\label{fig:ccf} }

\figcaption[too_ccpd.ps]
{ CCPD from the FR/RSS Monte Carlo simulations for the SAX97 observation. 
 The solid and dashed lines refer to DCF and MMD results, respectively.
\label{tooccpd} }

\figcaption[pv_ccpd.ps]
{ Same as Figure 6 but for the SAX96 observation. 
\label{pvccpd} } 

\figcaption[asca_ccpd.ps]
{ Same as Figure 6 but for the ASCA94 observation.
\label{ascaccpd} } 

\figcaption[comparison.ps]
{ Comparison of the lags derived directly from the DCF/MMD methods with
those estimated from the FR/RSS Monte Carlo simulations. 
\label{bijiao} }

\figcaption[cor_me2.ps,cor_fluxr.ps,cor_exosat.ps] { (a) The time lags
(0.1$-$1.5 keV vs 3.5$-$10 keV) are plotted against the fluxes in the
3.5$-$10 keV band; (b) the time lags (same as (a)) versus the ratios
of the maximum to the minimum fluxes in the 3.5$-$10 keV band;
(c) (logarithm of) the lags (0.1$-$2 keV vs 3$-$6 keV) with respect to
the the fluxes in the 3.5$-$10 keV band (the upper limit to the lag
for the EXOSAT 1985 Oct 24 \lc\ is taken from Tagliaferri \etal
1991). The dotted line refers to the relation predicted by the
homogeneous synchrotron model. In order to show individual errors on
the lags a small shift is applied to the flux values.
\label{correlation} }

\figcaption[cor_f_rms.ps,cor_f_t2.ps] { (a) Correlation between
$F_{var}$ and flux in the 1.5$-$10 keV band.  The value of $F_{var}$
during the flare of the ASCA94 is reported (see the text for details);
(b) behavior of the ``doubling time'' ($T_{2}$) versus the flux in
the 1.5$-$10 keV band.
\label{correlation2} }

\clearpage
\begin{deluxetable}{cccrc}
\tablecolumns{5}
\tablewidth{0pt}
\scriptsize
\tablecaption{Observation Log  \label{tab:log}}
\tablehead{
\colhead{Satellite} & \colhead{Instrument} & \colhead{Observing Time (UT)}
    & \colhead{Exposure} & \colhead{Observing Efficiency} \nl 
\colhead{} & \colhead{} & \colhead{(y m d h:m)} &  \colhead{(ks)} &
        \colhead{($\%$)} 
}
\startdata
\sax& LECS & 1997 Nov 22 16:03$-$ Nov 24 01:35 & 22 & 22 \nl
\sax& MECS & 1997 Nov 22 16:03$-$ Nov 24 01:35 & 63 & 53 \nl
\sax& LECS & 1996 Nov 20 00:16$-$ Nov 22 13:30 & 36 & 20 \nl
\sax& MECS & 1996 Nov 20 00:16$-$ Nov 22 13:30 & 108 & 52 \nl
ASCA\tablenotemark{a} & GIS  & 1994 May 19 04:30$-$May 21 07:55 & 100& 
                                                           55\nl  
\enddata
\tablenotetext{a}{Only GIS data are analyzed}
\end{deluxetable}
\clearpage
\begin{deluxetable}{rccccccrr}
\tablecolumns{9}
\scriptsize
\tablecaption{Variability Parameters \tablenotemark{a} \label{tab:var}}
\tablehead{
\colhead{Band} & \colhead{Flux \tablenotemark{b}} &
\colhead{$\overline{F}$} & \colhead{$\sigma_{F}$} &
\colhead{$\Delta_{F}$} & \colhead{$\sigma_{exc}$} &
\colhead{$F_{var}$} & \colhead{$\chi^{2}$(dof) \tablenotemark{c}} &
          \colhead{$T_{2}$} \nl 
\colhead{(keV)} & \colhead{} & \colhead{(cts ${\rm s^{-1}}$)} & 
    \colhead{(cts ${\rm s^{-1}}$)} & \colhead{(cts ${\rm s^{-1}}$)} &
    \colhead{(cts ${\rm s^{-1}}$)} & \colhead{} & \colhead{} &
    \colhead{(hours)} \nl
\cline{1-9} \nl
\multicolumn{9}{c}{SAX97}
}
\startdata
   0.1-1.5 &16.6 &1.56 &0.36 &0.08 &0.35 &0.22 &1711(86)  &3.37$\pm$0.61 
\nl
   1.5-3.5 &6.78 &1.00 &0.28 &0.06 &0.27 &0.27 &4270(215) &1.94$\pm$0.29
\nl
   3.5-10  &4.02 &0.40 &0.13 &0.04 &0.12 &0.30 &2171(215) &1.81$\pm$0.31
\nl
   1.5-10  &10.8 &1.41 &0.40 &0.08 &0.39 &0.28 &6192(215) &2.04$\pm$0.28
\nl
\cutinhead {SAX96}
   0.1-1.5 &11.9 &1.01 &0.12 &0.07 &0.10 &0.10 &462(141)  &22.11$\pm$4.16
\nl
   1.5-3.5 &4.46 &0.87 &0.12 &0.06 &0.11 &0.13 &1811(383) &14.46$\pm$2.59
\nl
   3.5-10  &2.88 &0.41 &0.07 &0.04 &0.06 &0.13 &1156(383) &8.17$\pm$1.61
\nl
   1.5-10  &7.34 &1.28 &0.18 &0.07 &0.16 &0.13 &2509(383) &9.96$\pm$1.88
\nl
\cutinhead {ASCA94}
   0.5-1.5 &11.9 &1.51 &0.34 &0.08 &0.33 &0.22 &5933(330)  &5.61$\pm$1.08
\nl
   1.5-3.5 &5.40 &1.42 &0.47 &0.08 &0.46 &0.33 &11981(330) &4.83$\pm$0.93
\nl
   3.5-10  &3.76 &0.38 &0.17 &0.04 &0.16 &0.43 &5824(330)  &4.53$\pm$0.89
\nl
   1.5-10  &9.16 &1.81 &0.63 &0.09 &0.62 &0.35 &17191(330) &4.89$\pm$0.67
\nl
\enddata 
\tablenotetext{a}{All parameters in this table refer to a 300s
   binning. See the text for the definition of each parameter}
\tablenotetext{b}{Mean flux in unit of $10^{-11}$ erg ${\rm cm^{-2}
    s^{-1}}$}
\tablenotetext{c}{Constant fits}
\end{deluxetable}
\clearpage
\begin{deluxetable}{lcc}
\tablecolumns{3}
\tablewidth{0pc}
\tablecaption{PDS Parameters \label{tab:pds}}
\tablehead{
\colhead{Observation} &
  \colhead{Fitting Slope \tablenotemark{a}} &
        \colhead{$\chi^{2}(dof)$ \tablenotemark{a}}  \nl
 }
\startdata
SAX97  & 2.17$\pm$0.10, 1.99$\pm$0.10 & 3.4(8), 9.0(8) \nl
SAX96  & 1.54$\pm$0.07, 1.40$\pm$0.07 & 12(9), 13(9)   \nl
ASCA94 & 2.19$\pm$0.23, 2.15$\pm$0.49 & 1.1(5), 0.9(3) \nl
\enddata
\tablenotetext{a}{The fitting region is from $\sim 6 \times 10^{-5}$ to
    $\sim 6 \times 10^{-3}$ Hz (SAX97), $\sim 8 \times 10^{-3}$ Hz (SAX96)
    and $\sim 1.5 \times 10^{-3}$ Hz (ASCA94); the second values in
    this column refer to the PDS after the removal of a linear trend}
\end{deluxetable}
\clearpage
\begin{deluxetable}{cccccccccc}
\tablecolumns{10}
\tablewidth{0pt}
\scriptsize
\tablecaption{Cross Correlations and Lags: DCF and MMD Results  
         \label{tab:lag}}
\tablehead{
\colhead{First} & \colhead{Second} & &
   \multicolumn{2}{c}{DCF} & & \multicolumn{2}{c}{MMD} & &
   \colhead{$r_{0}$}  \nl
\cline{4-5} \cline{7-8} \nl
\colhead{Series}  & \colhead{Series} & &  
     \colhead{Lag} & \colhead{90\% C.L.} & & 
     \colhead{Lag} & \colhead{90\% C.L.} & &  \nl 
\cline{1-2} \cline{4-5} \cline{7-8}  \nl
\multicolumn{2}{c}{(keV)} & & 
  \multicolumn{2}{c}{(hours)} & & \multicolumn{2}{c}{(hours)} & & \nl
\cline{1-10} \nl
\multicolumn{10}{c}{SAX97}
}
\startdata
0.1$-$1.5 & 1.5$-$3.5 && 0.30$\pm$0.07 & 0.19,0.41 &&
          0.17$\pm$0.05 & 0.08,0.26 && 0.92 \nl
1.5$-$3.5 & 3.5$-$10  && 0.00$\pm$0.03 & -0.05,0.05 &&
          0.06$\pm$0.03 & 0.01,0.12 && 0.91  \nl  
0.1$-$1.5 & 3.5$-$10  && 0.46$\pm$0.07 & 0.34,0.58 &&
          0.30$\pm$0.06 & 0.20,0.41  && 0.87  \nl
\cutinhead {SAX96}
0.1$-$1.5 & 1.5$-$3.5 && 1.96$\pm$0.15 & 1.71,2.21 && 
          1.01$\pm$0.22 & 0.63,1.41 && 0.73  \nl
1.5$-$3.5 & 3.5$-$10  && 1.91$\pm$0.10 & 1.74,2.08 &&
          1.72$\pm$0.18 & 1.40,2.05 && 0.63  \nl
0.1$-$1.5 & 3.5$-$10  && 4.09$\pm$0.16 & 3.83,4.35 &&
          4.08$\pm$0.36 & 3.48,4.69 && 0.52  \nl
\cutinhead {ASCA94}
0.5$-$1.5 & 1.5$-$3.5 && 0.29$\pm$0.27 & -0.16,0.77 && 
           0.30$\pm$0.06 & 0.21,0.39 && 0.96 \nl
1.5$-$3.5 & 3.5$-$10  && 0.57$\pm$0.30 & 0.09,1.08 && 
           0.50$\pm$0.06 & 0.39,0.60 && 0.95  \nl
0.5$-$1.5 & 3.5$-$10  && 0.88$\pm$0.30 & 0.40,1.38 && 
           0.85$\pm$0.07 & 0.73,0.97 && 0.93  \nl
\enddata
\end{deluxetable}
\clearpage
\begin{deluxetable}{cccccccccc}
\tablecolumns{10}
\tablewidth{0pt}
\scriptsize
\tablewidth{0pc}
\tablecaption{Results of the FR/RSS Simulations \label{tab:mcs}}
\tablehead{
\colhead{First} & \colhead{Second} & \colhead{} & \multicolumn{3}{c}{DCF} 
& 
    \colhead{} & \multicolumn{3}{c}{MMD} \nl 
\cline{4-6}  \cline{8-10} \nl
\colhead{Series} & \colhead{Series} & & \colhead{Lag} & 
    \colhead{68\% C.L.}
    & \colhead{90\% C.L.} & \colhead{} & \colhead{Lag} & \colhead{68\% 
     C.L.} & \colhead{90\% C.L.} \\
\cline{1-2} \cline{4-6} \cline{8-10} \\
\multicolumn{2}{c}{(keV)} & &\multicolumn{3}{c}{(hours)} & 
     \colhead{} 
    & \multicolumn{3}{c}{(hours)} \\
\cline{1-10} \\
\multicolumn{10}{c}{SAX97}
 } 
\startdata
0.1$-$1.5 & 1.5$-$3.5 && 0.14 & -0.10,0.40 & -0.28,0.57 &
                      & 0.16 & 0.04,0.28 & -0.05,0.36 \nl
1.5$-$3.5 & 3.5$-$10 && 0.07 & -0.10,0.25 & -0.22,0.35 &
                       & 0.06 & -0.03,0.15 & -0.09,0.20 \nl
0.1$-$1.5 & 3.5$-$10  && 0.32 & 0.06,0.59 & -0.15,0.78 & 
                       & 0.31 & 0.16,0.45 & 0.05,0.55 \nl
\cutinhead {SAX96}
0.1$-$1.5 & 1.5$-$3.5 && 1.95 & 1.00,2.92 & 0.28,3.64 & 
                      & 1.18 & 0.39,1.96 & -0.01,2.84 \nl
1.5$-$3.5 & 3.5$-$10  && 2.14 & 1.38,2.89 & 0.87,3.43 & 
                      & 1.77 & 1.04,2.48 & 0.68,3.10 \nl
0.1$-$1.5 & 3.5$-$10  && 4.27 & 3.21,5.30 & 2.52,6.04 &
                      & 4.30 & 3.03,5.60 & 2.10,6.59 \nl
\cutinhead {ASCA94}
0.5$-$1.5 & 1.5$-$3.5 && 0.32 & -0.40,1.04 & -0.87,1.53 &   
                      & 0.31 & 0.12,0.51 & -0.02,0.63 \nl  
1.5$-$3.5 & 3.5$-$10  && 0.44 & -0.27,1.18 & -0.84,1.70 & 
                      & 0.52 & 0.24,0.80 & 0.14,0.99 \nl
0.5$-$1.5 & 3.5$-$10  && 0.85 & 0.15,1.59 & -0.25,2.10 &
                      & 0.88 & 0.58,1.18 & 0.39,1.40 \nl 
\enddata
\end{deluxetable}
\clearpage
\plotone{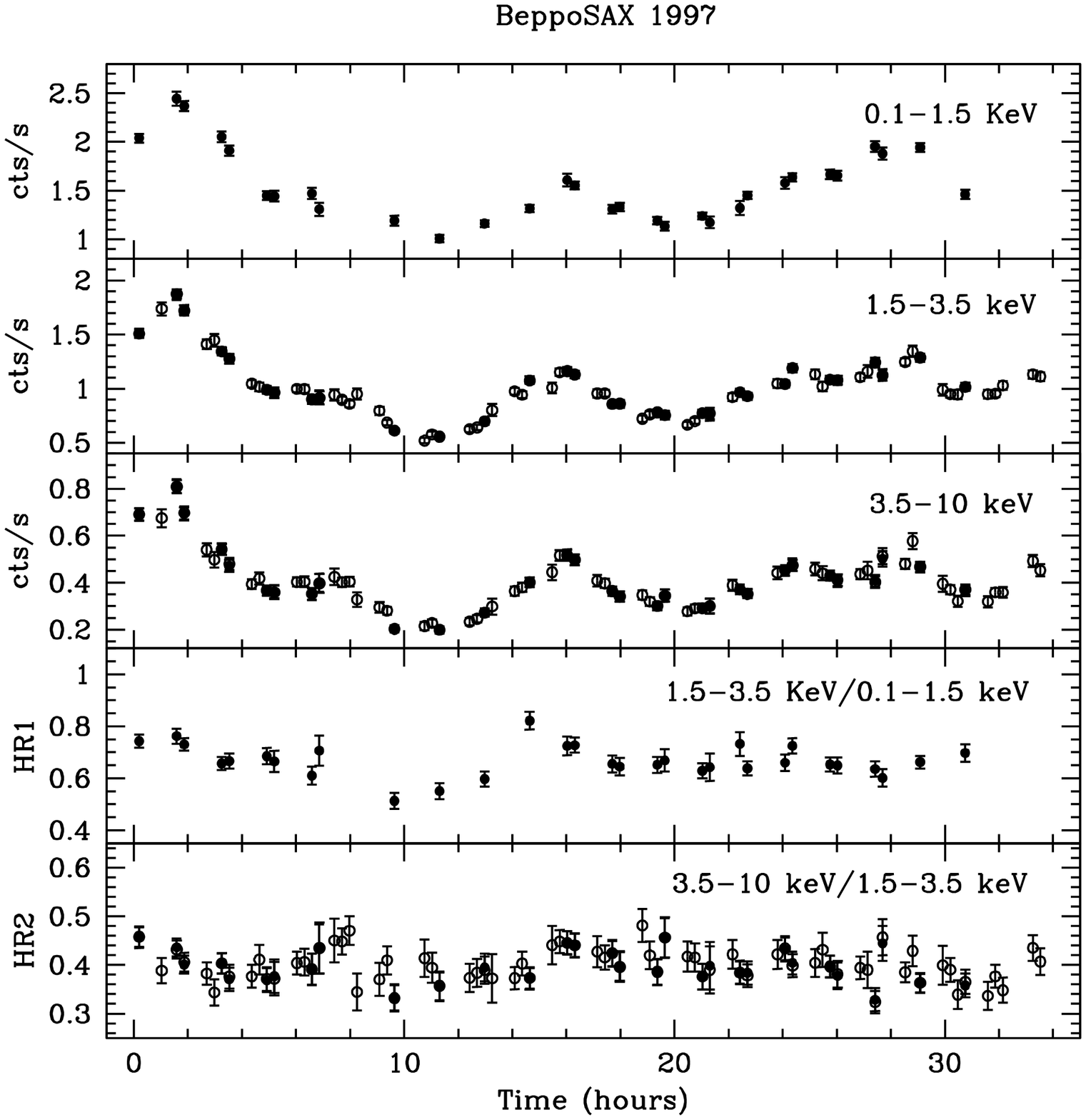}

\clearpage
\plotone{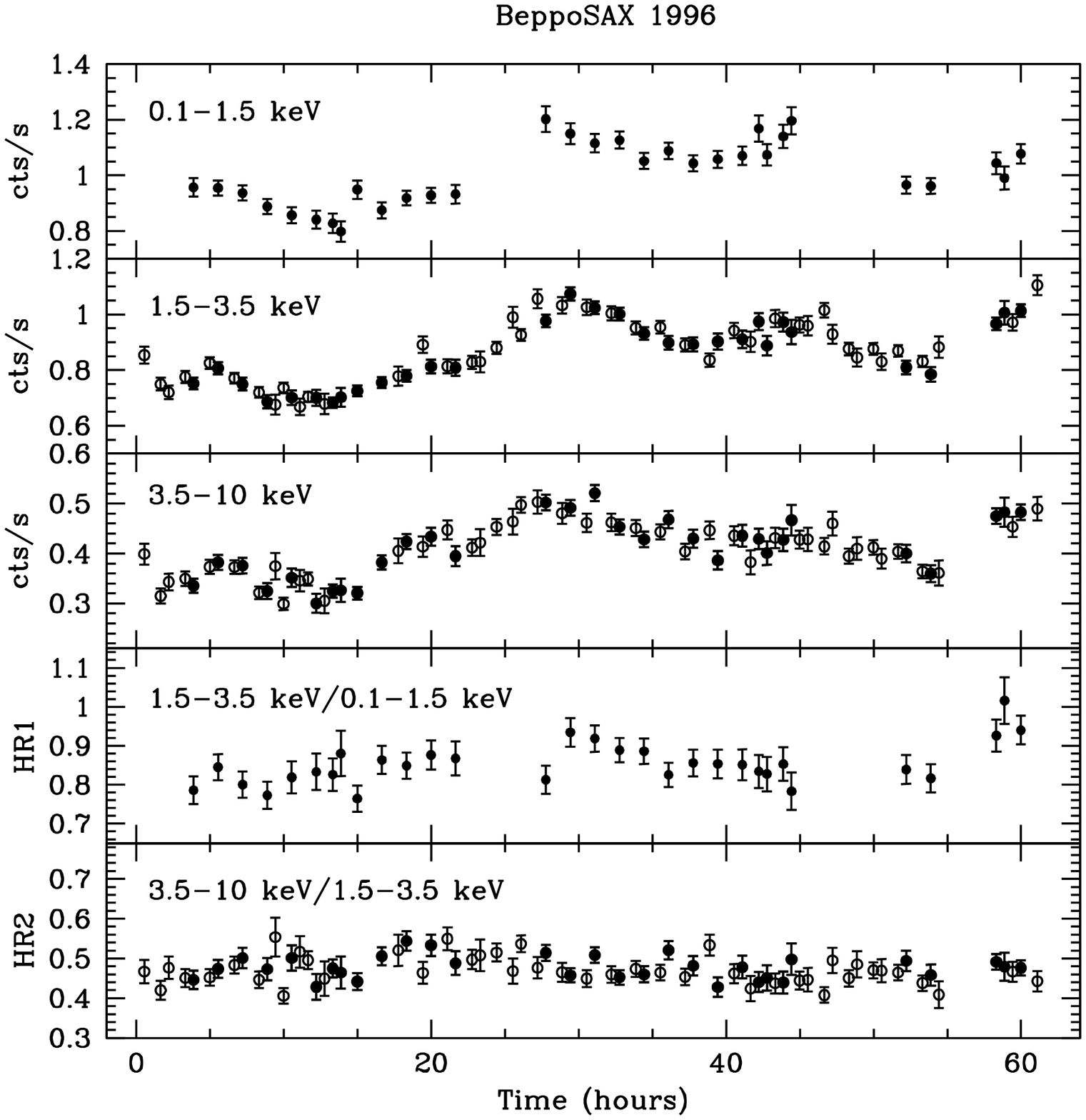}

\clearpage
\plotone{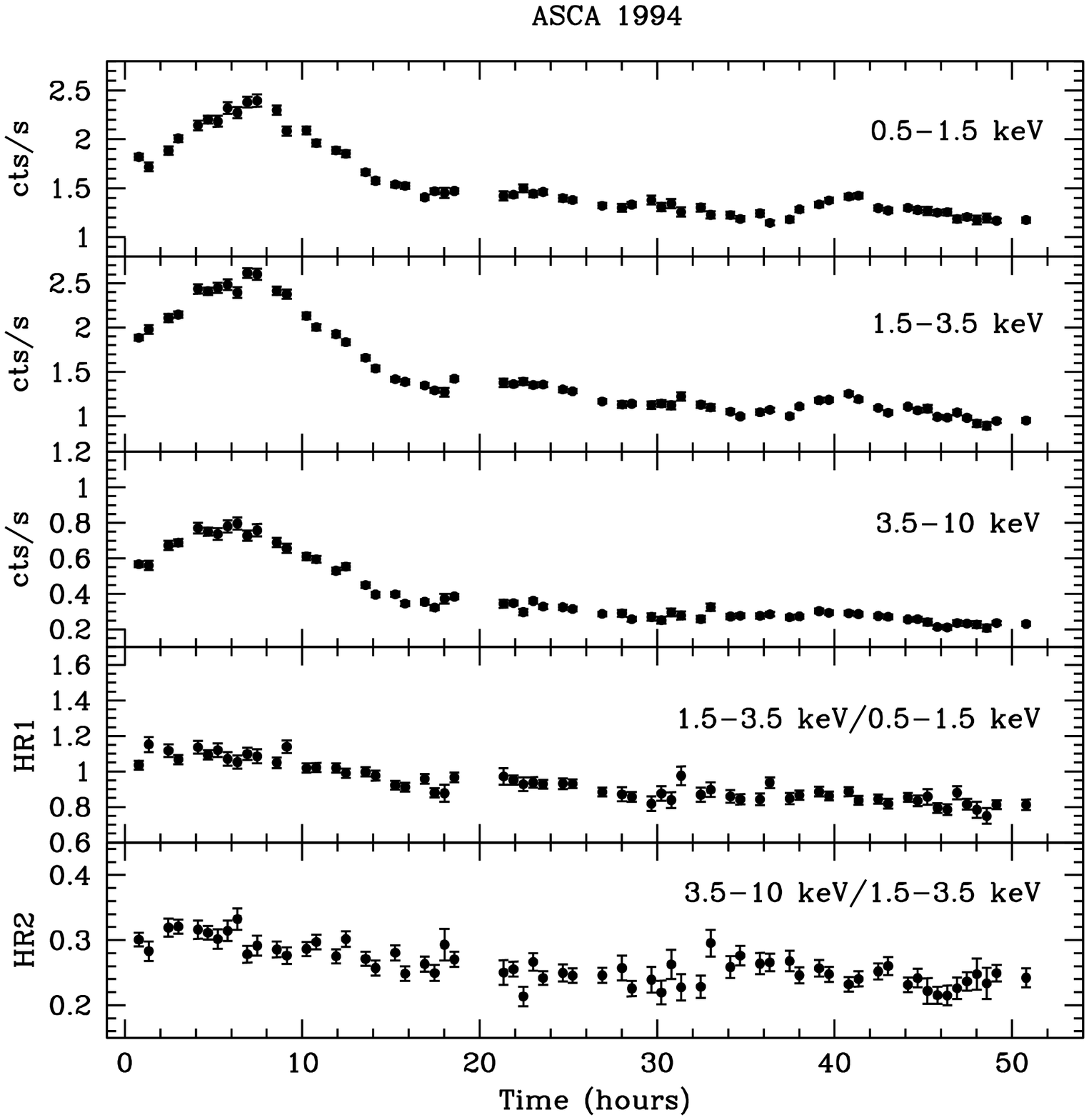}
\clearpage
\begin{figure}
\plotfiddle{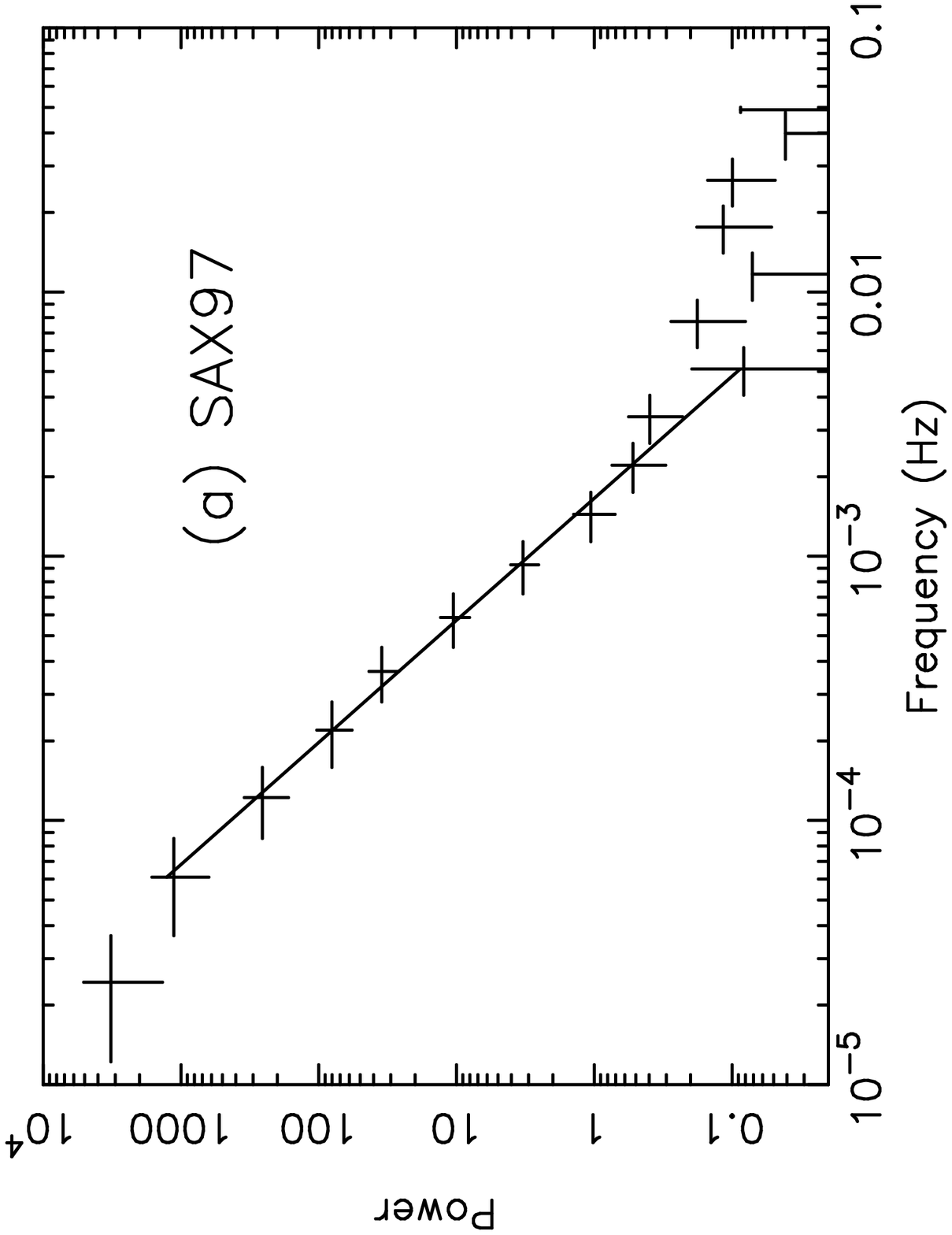}{0cm}{270}{40}{40}{-140}{320}
\plotfiddle{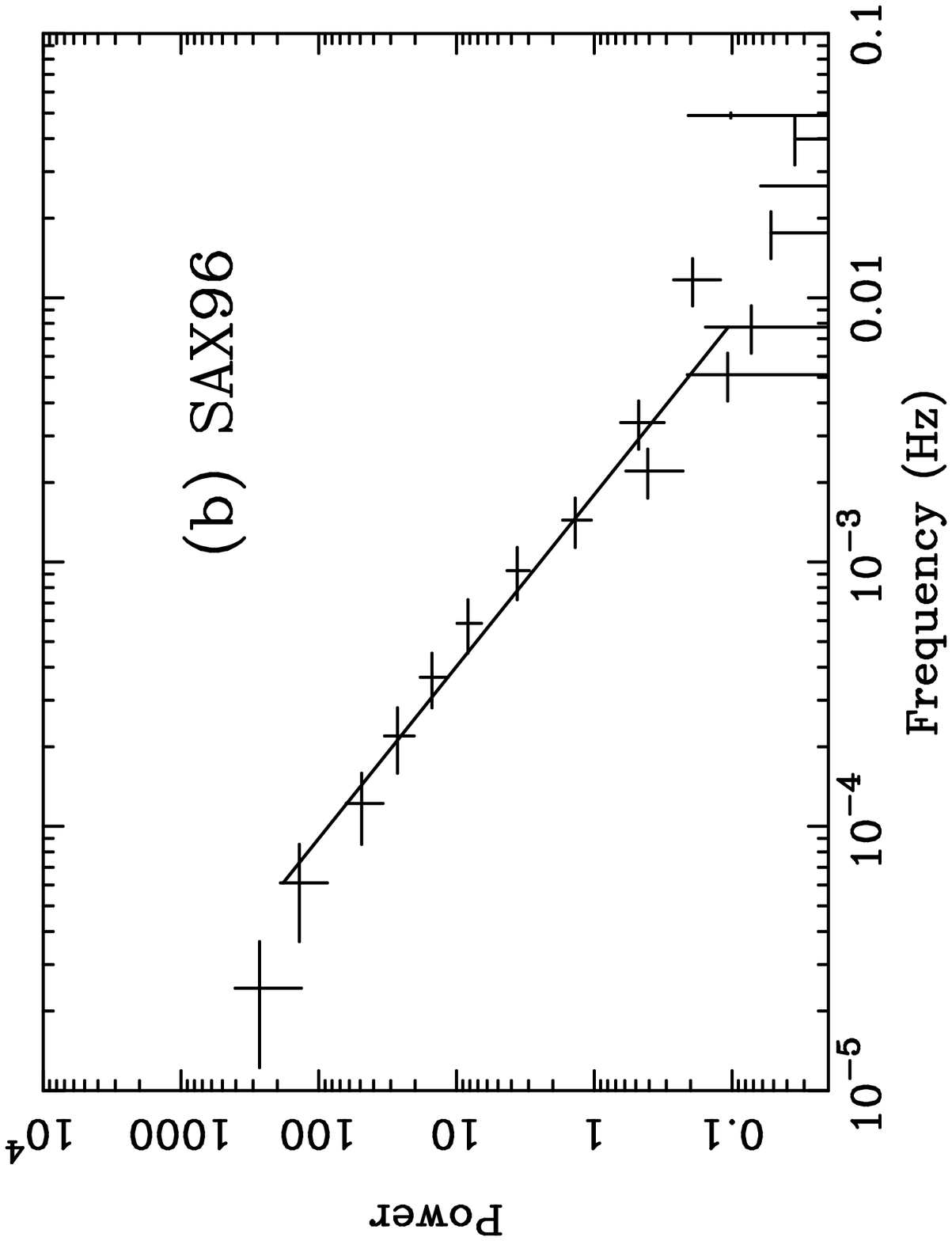}{0cm}{270}{40}{40}{-140}{180}
\plotfiddle{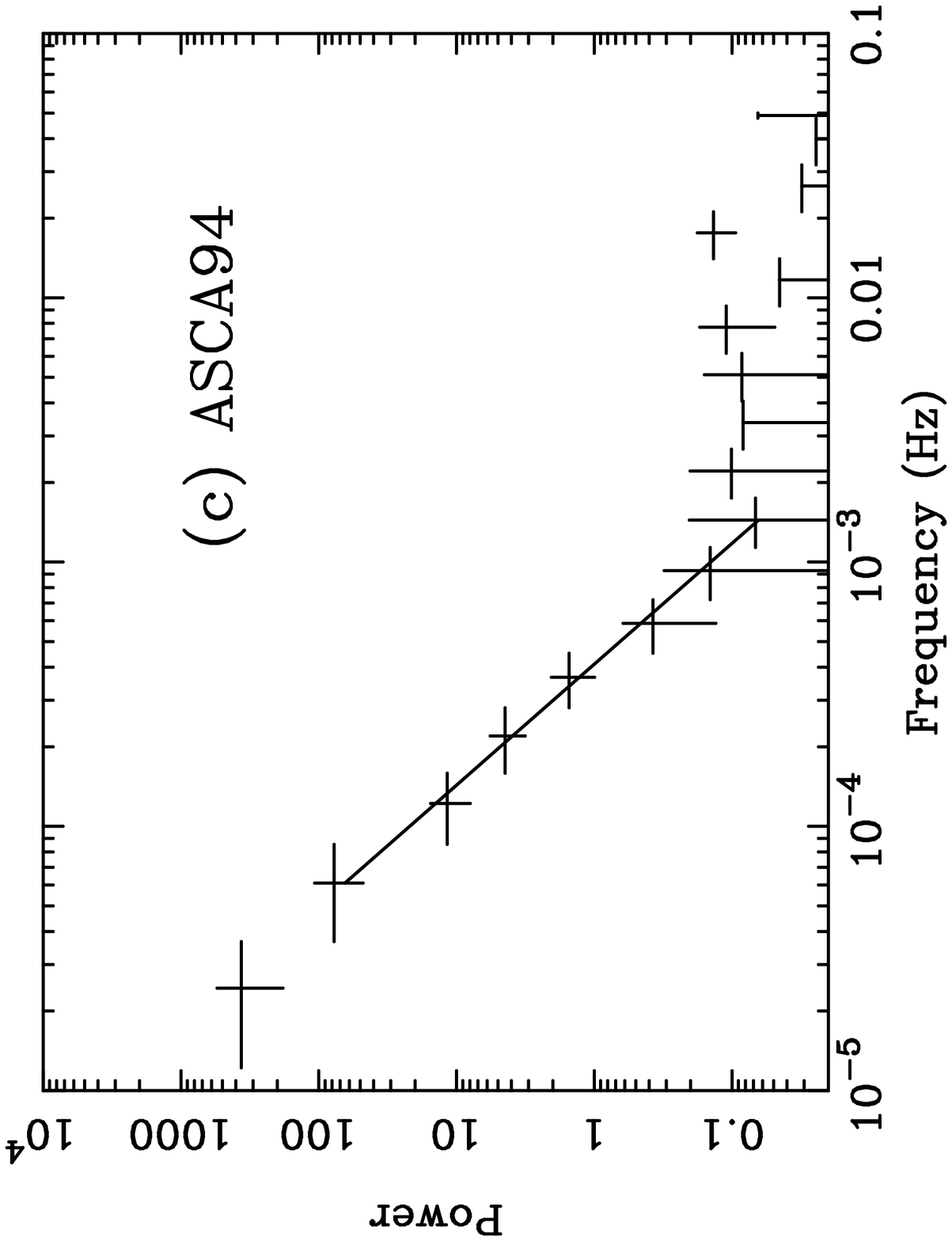}{0cm}{270}{40}{40}{-140}{40}
\end{figure}
\clearpage
\begin{figure}
\plotfiddle{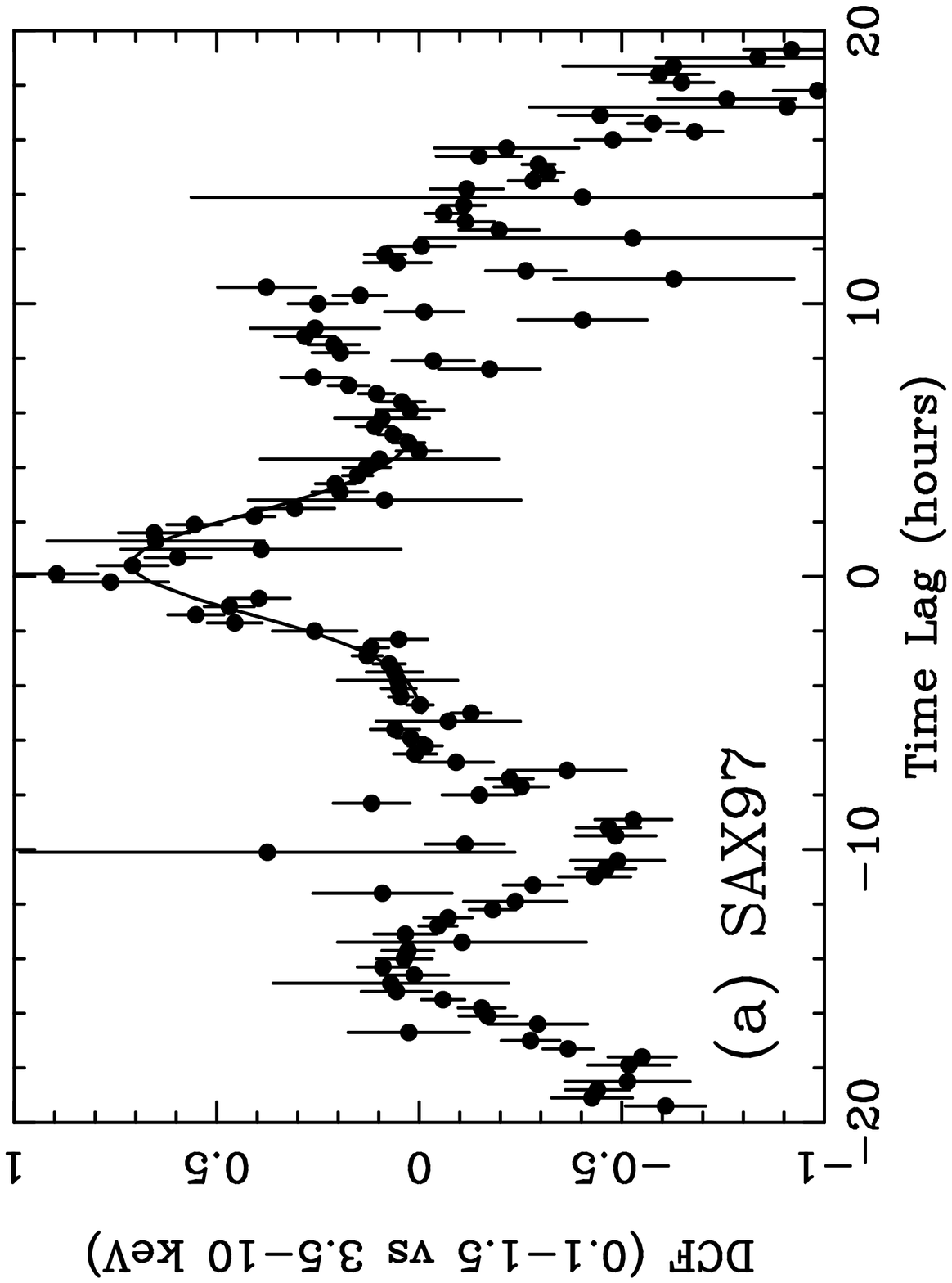}{0cm}{270}{40}{40}{-260}{220}
\plotfiddle{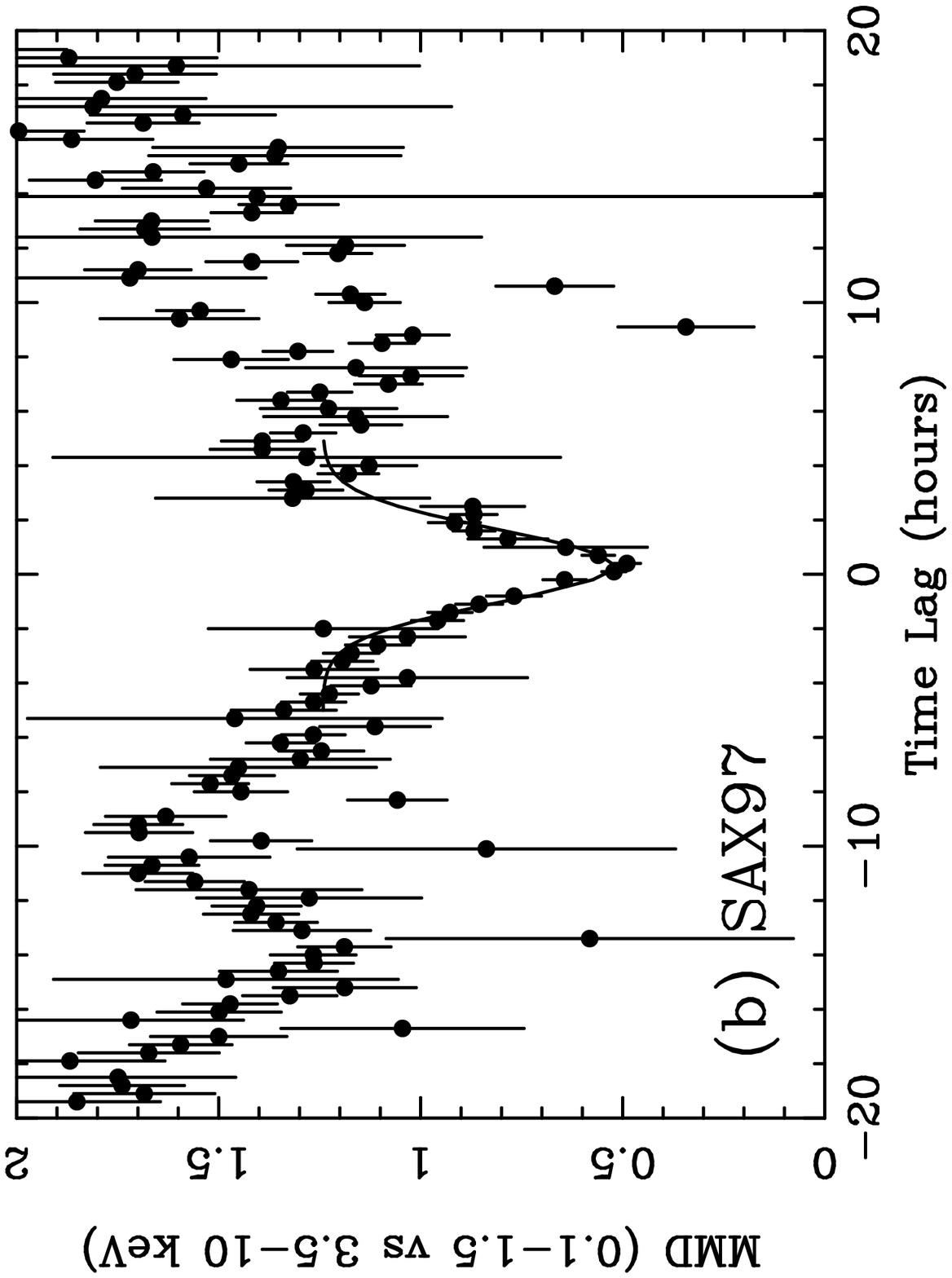}{0cm}{270}{40}{40}{-20}{265}
\plotfiddle{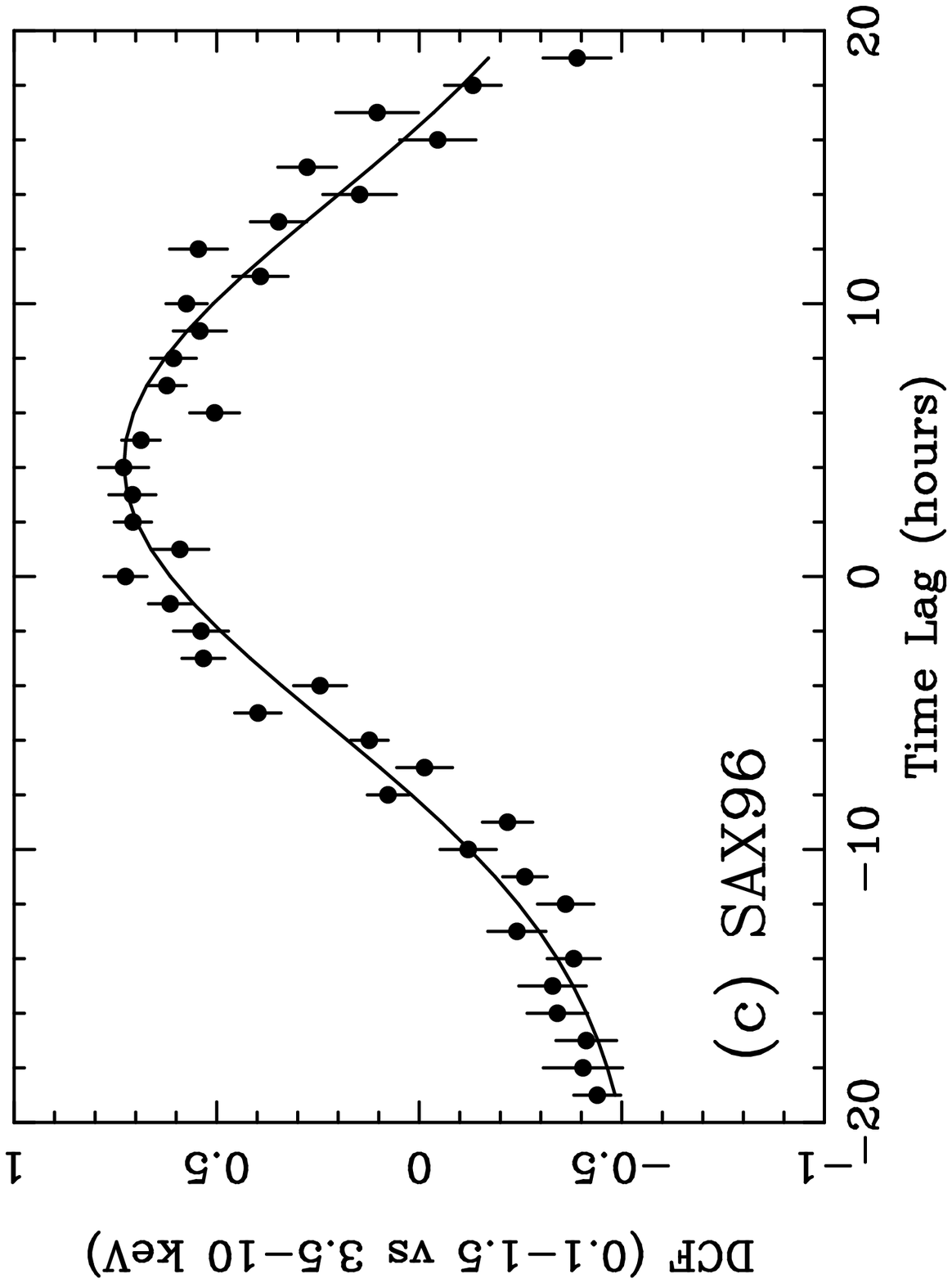}{0cm}{270}{40}{40}{-260}{140}
\plotfiddle{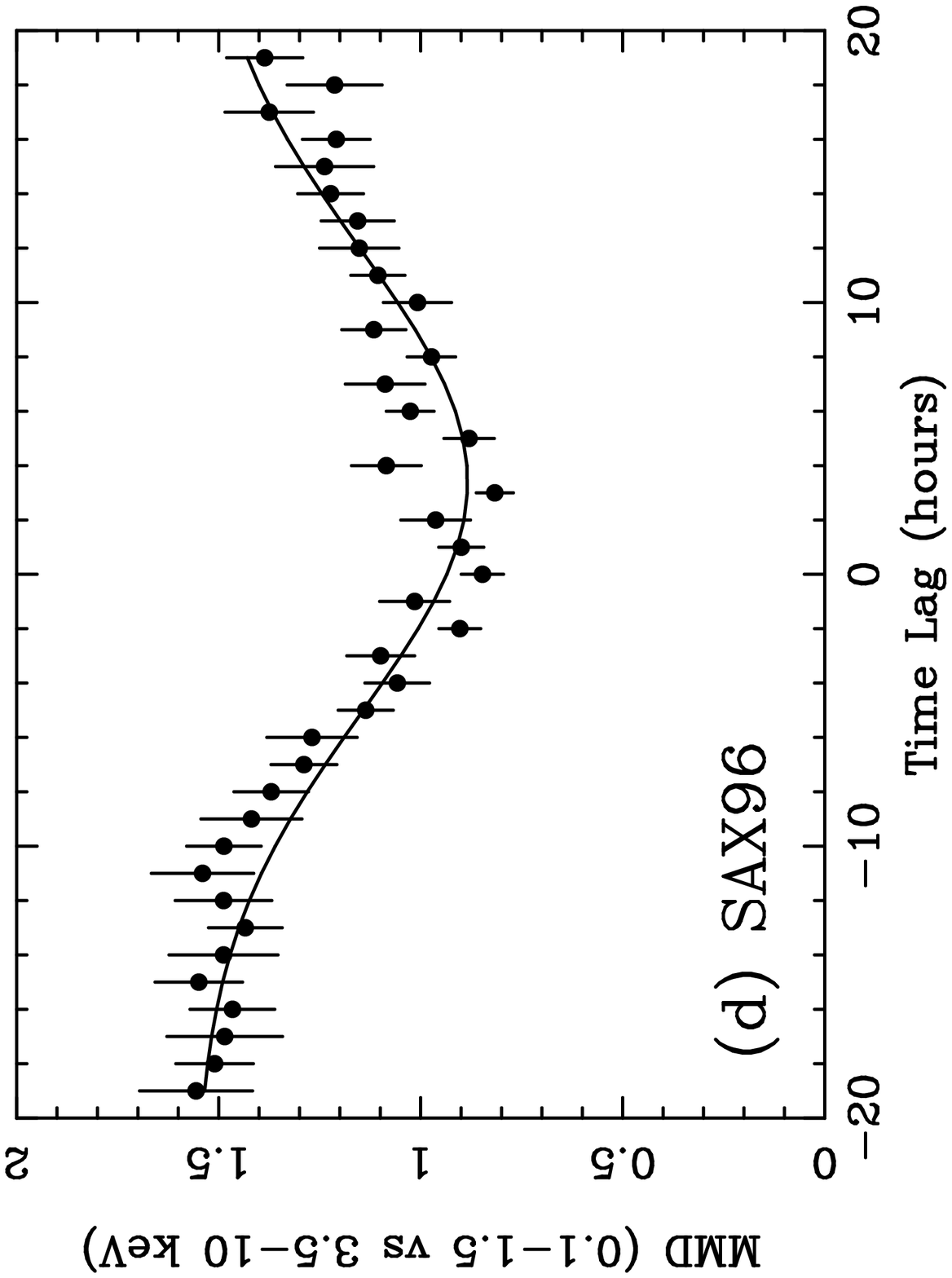}{0cm}{270}{40}{40}{-20}{185}
\plotfiddle{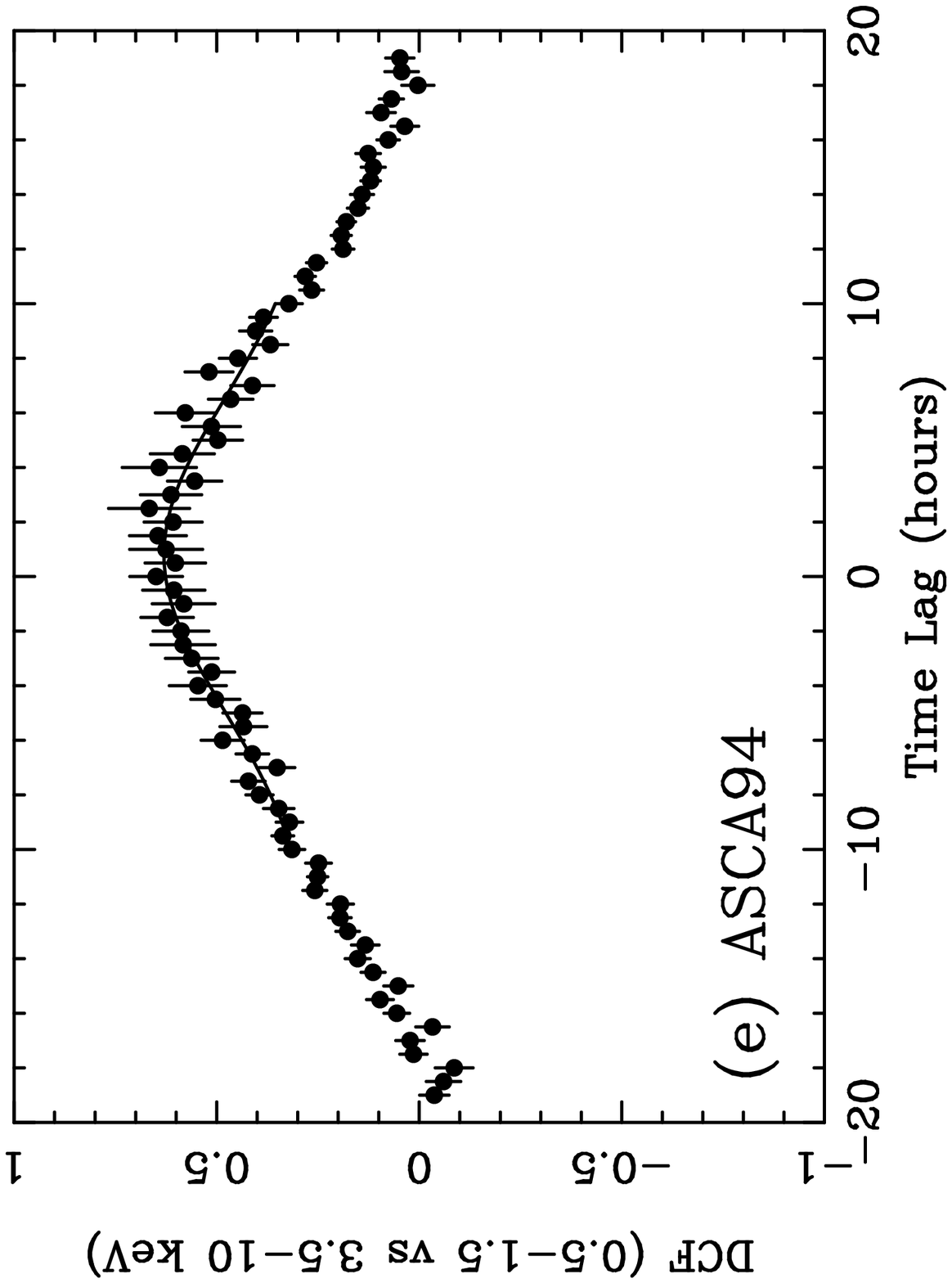}{0cm}{270}{40}{40}{-260}{60}
\plotfiddle{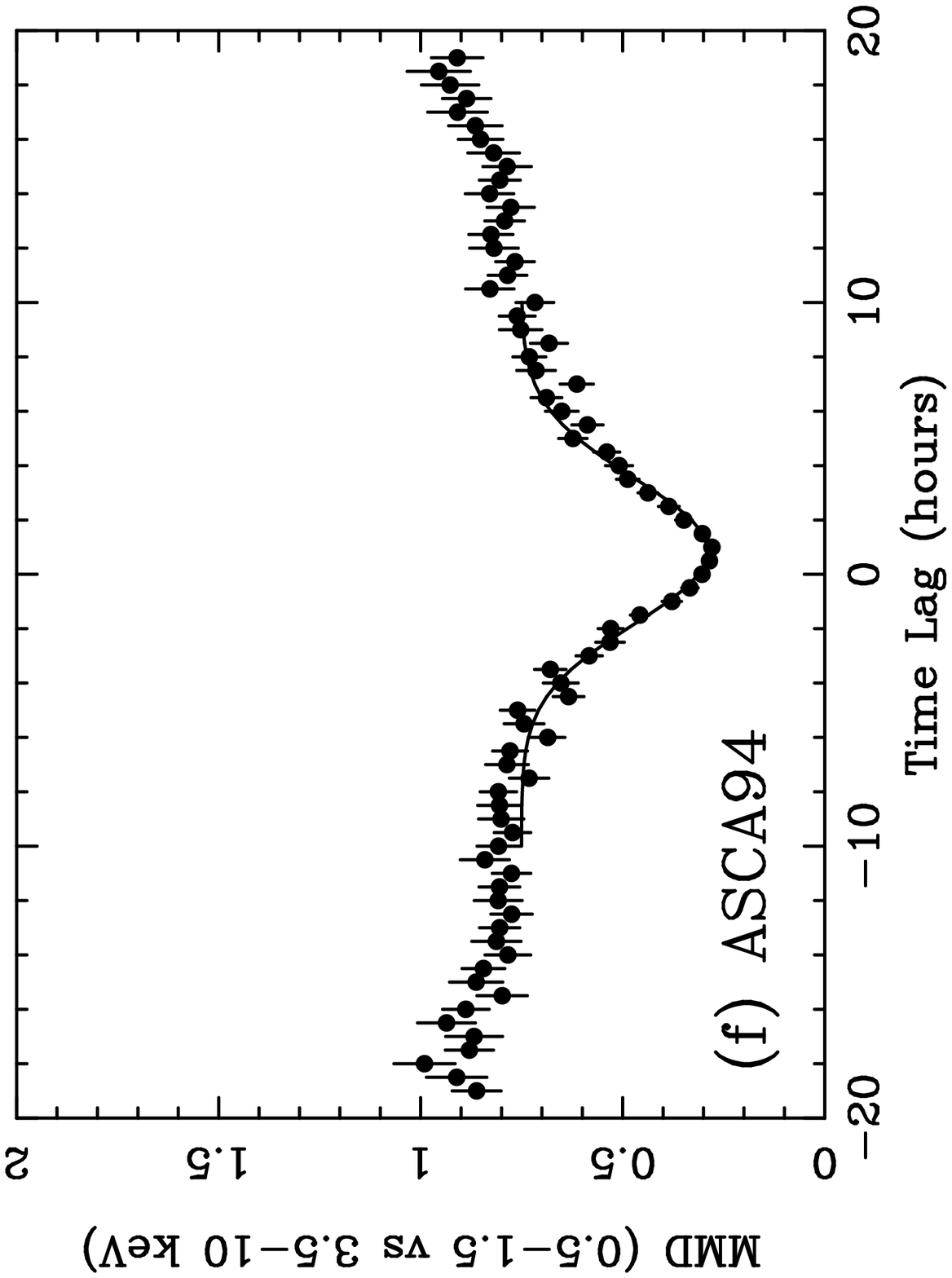}{0cm}{270}{40}{40}{-20}{105}
\end{figure}
\clearpage
\plotone{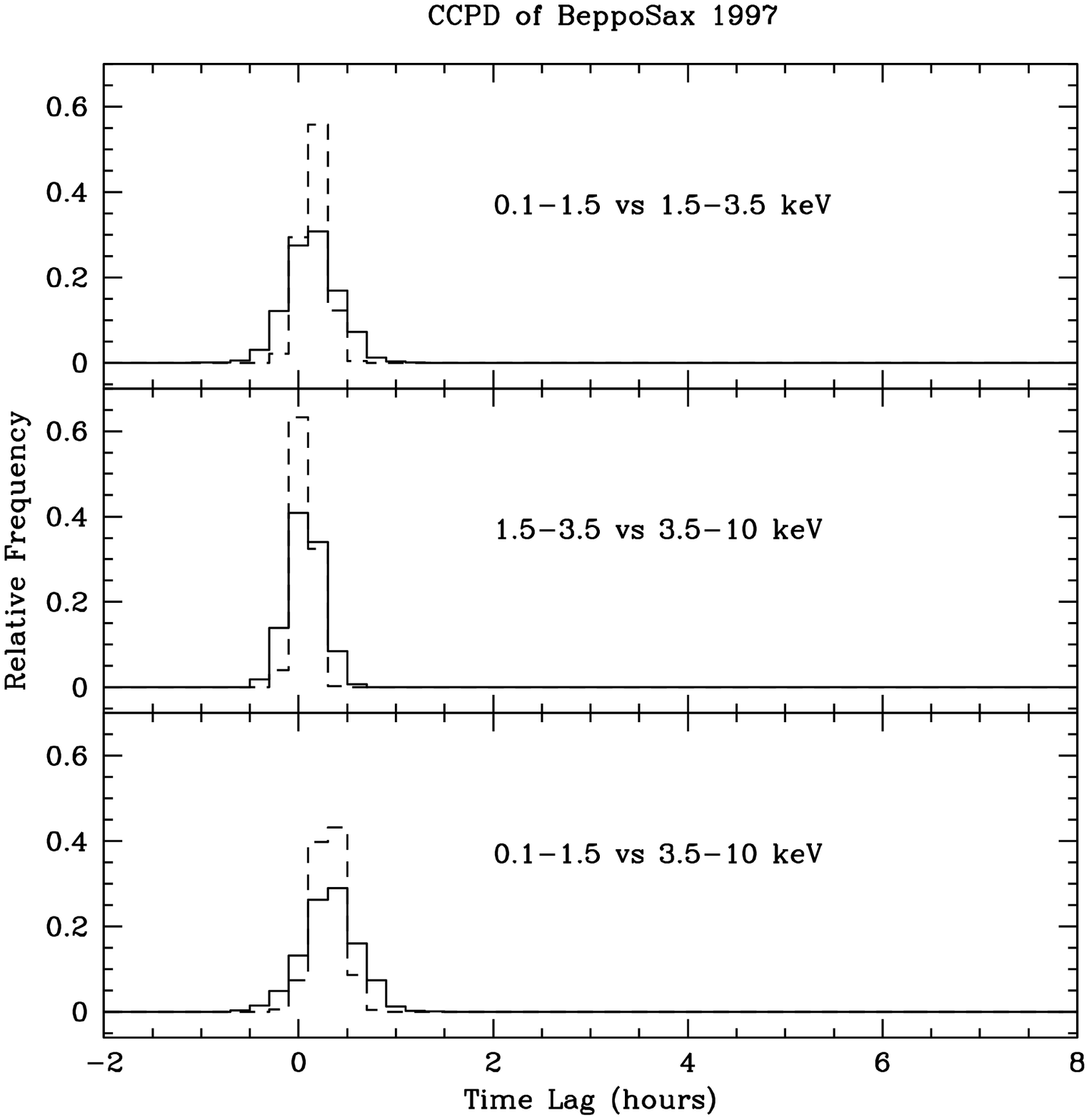}

\clearpage
\plotone{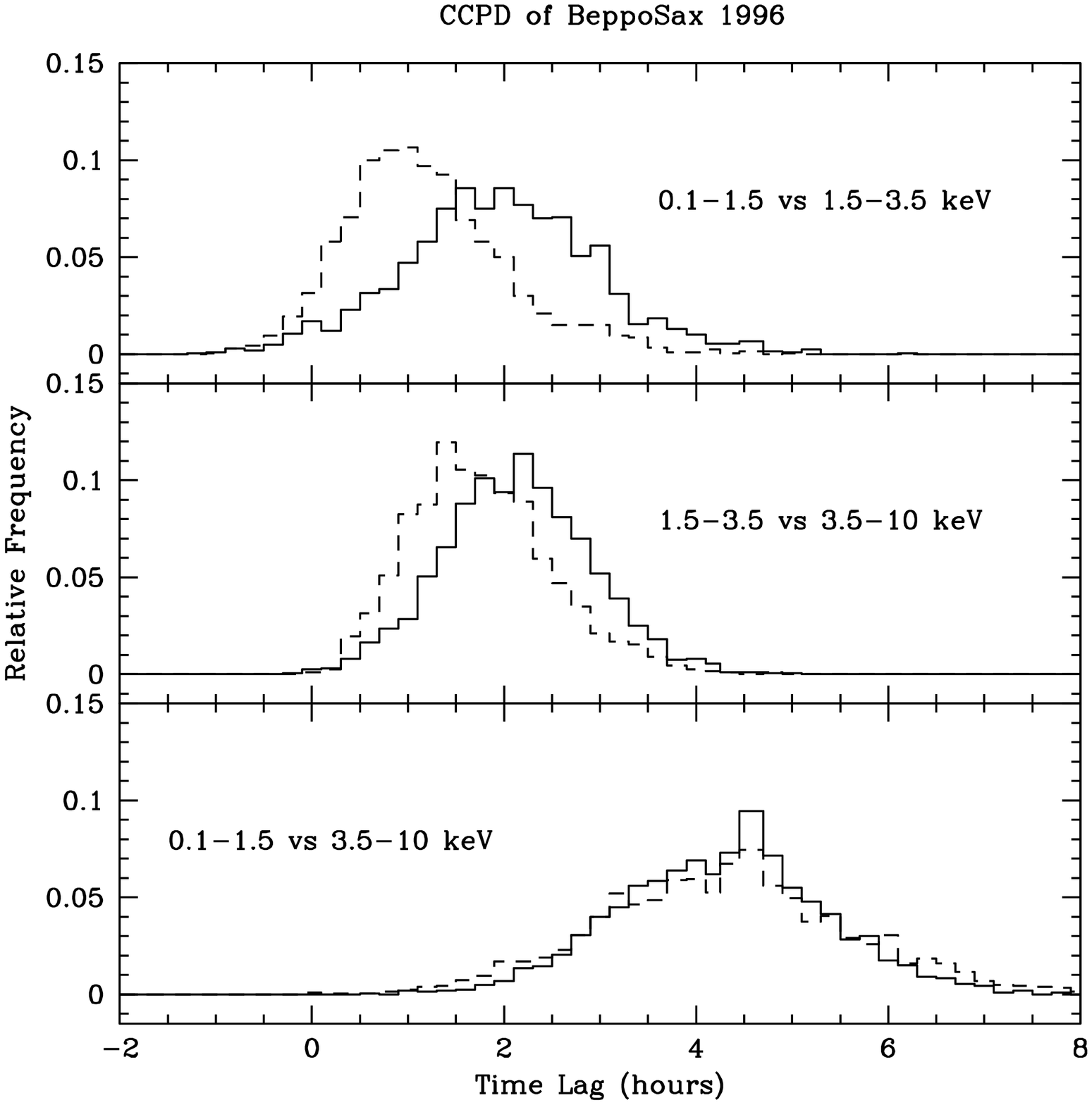}

\clearpage
\plotone{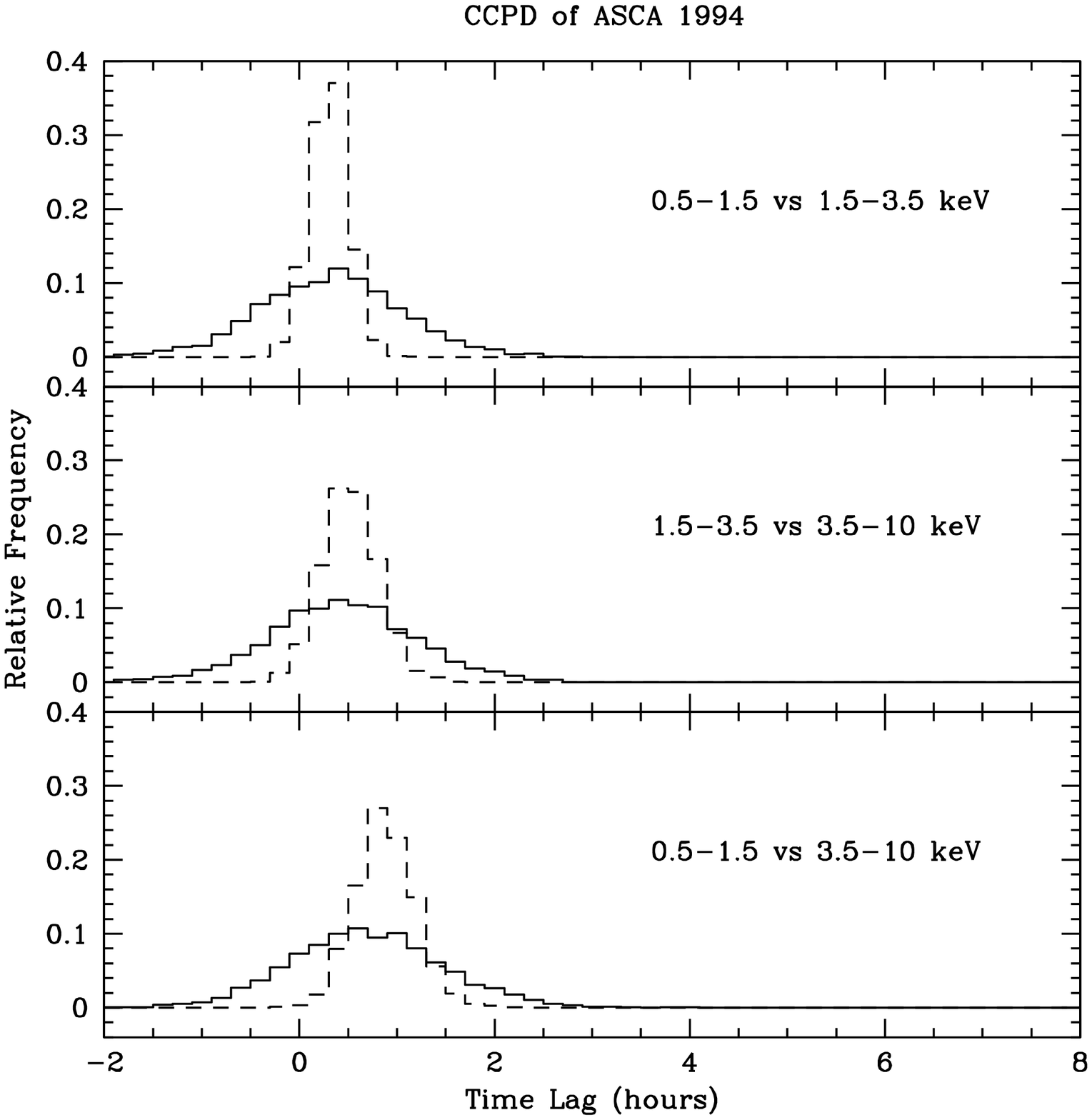}
\clearpage
\plotfiddle{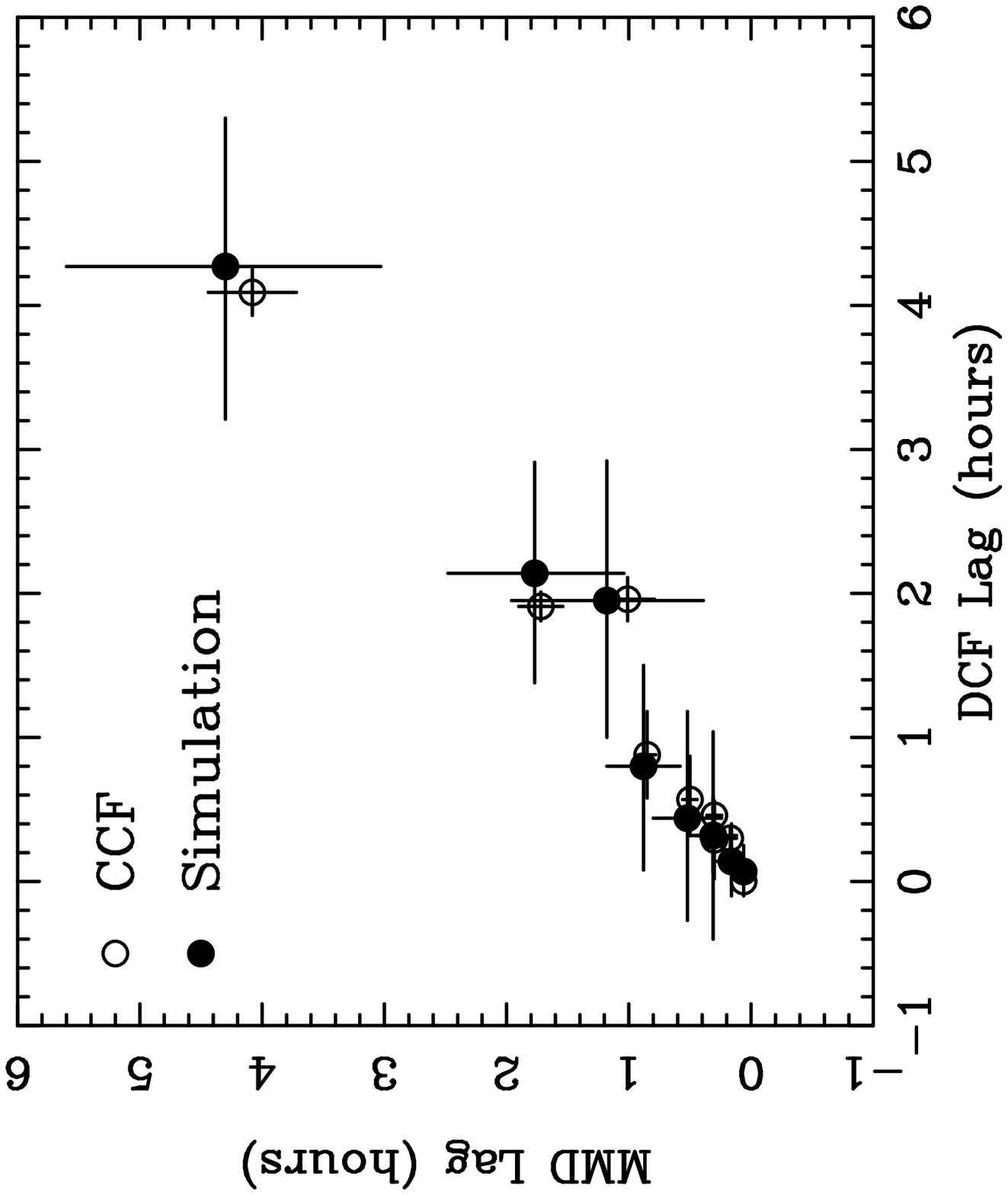}{0cm}{270}{50}{50}{-170}{0}
\clearpage
\begin{figure}
\plotfiddle{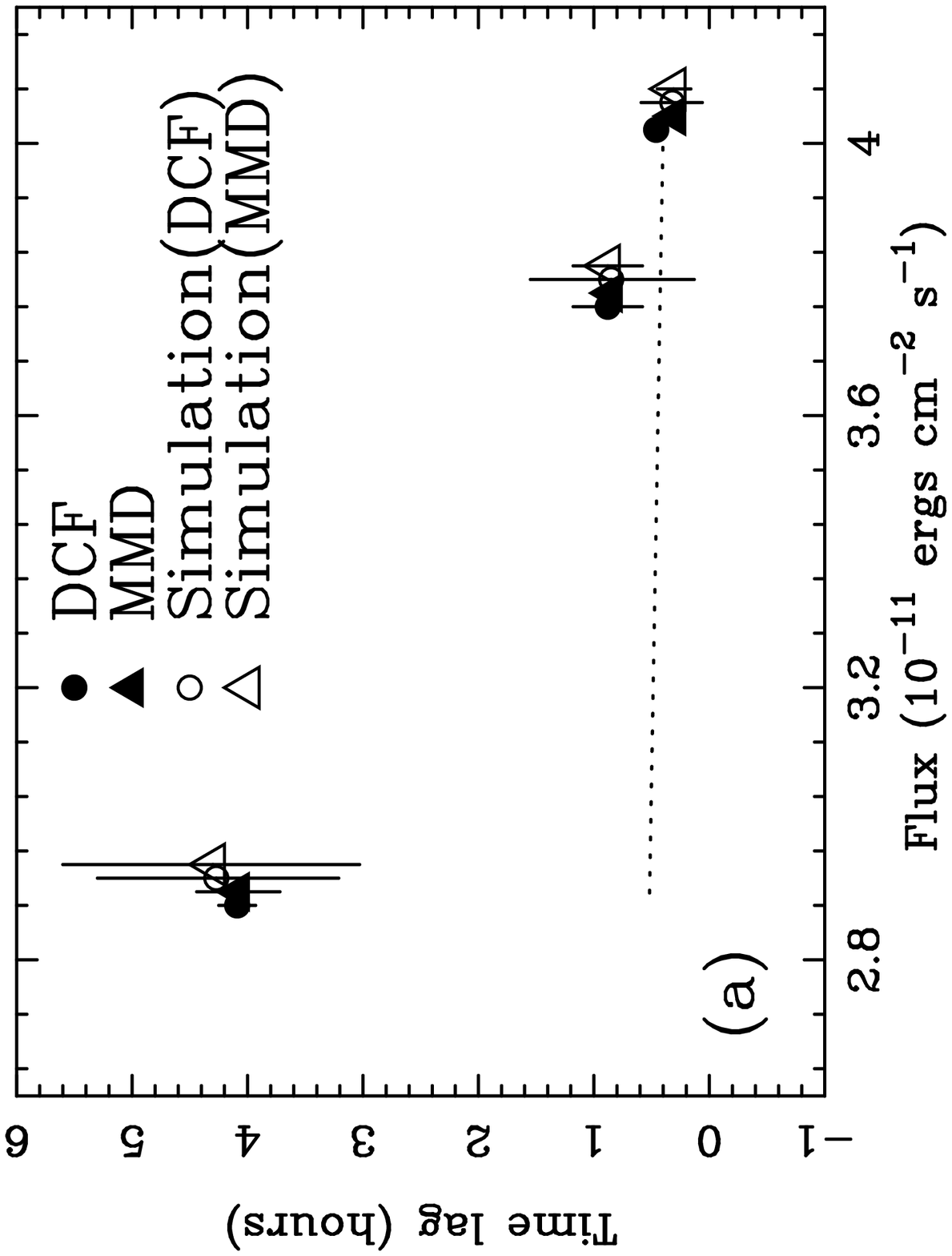}{0cm}{270}{40}{40}{-140}{320}
\plotfiddle{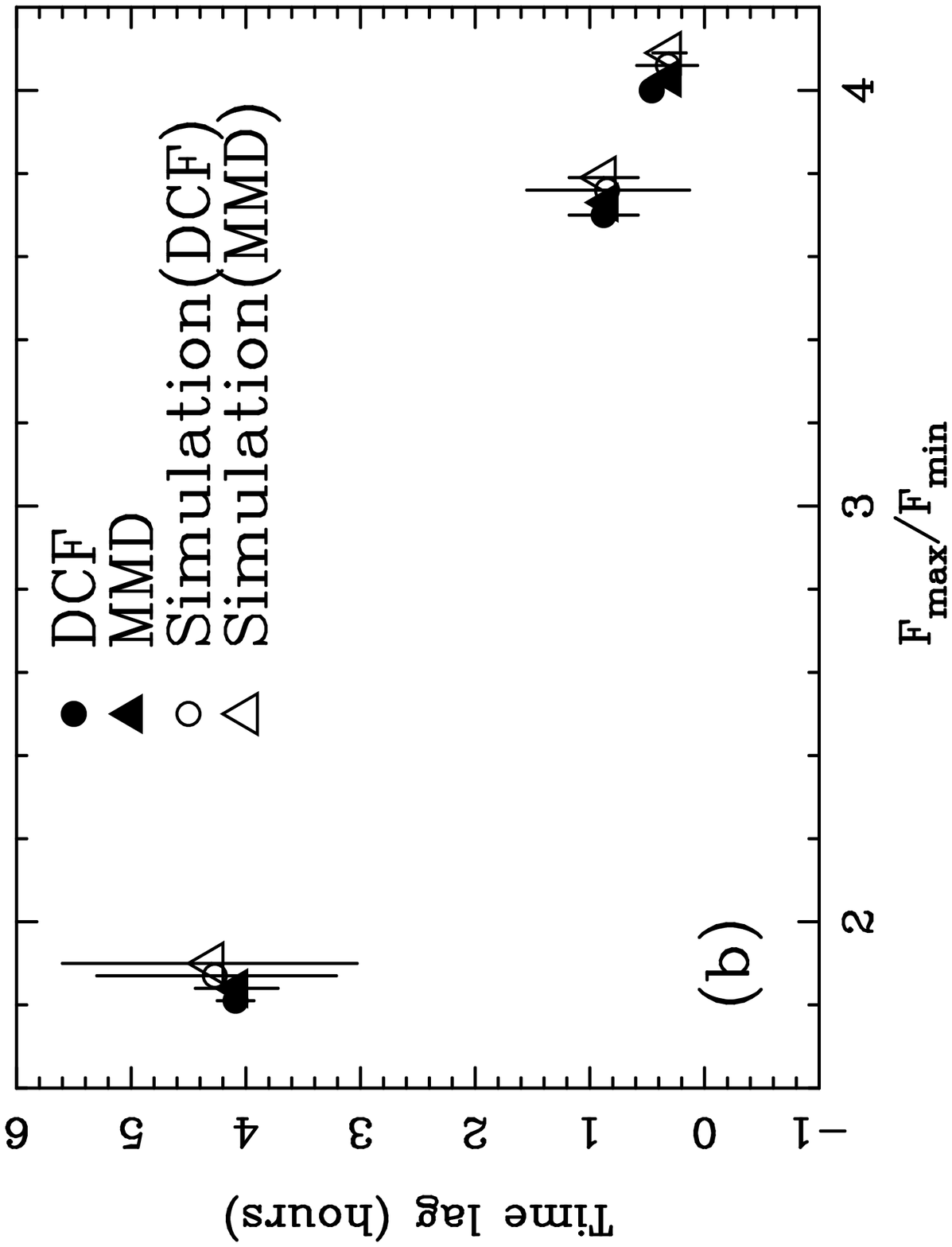}{0cm}{270}{40}{40}{-140}{180}
\plotfiddle{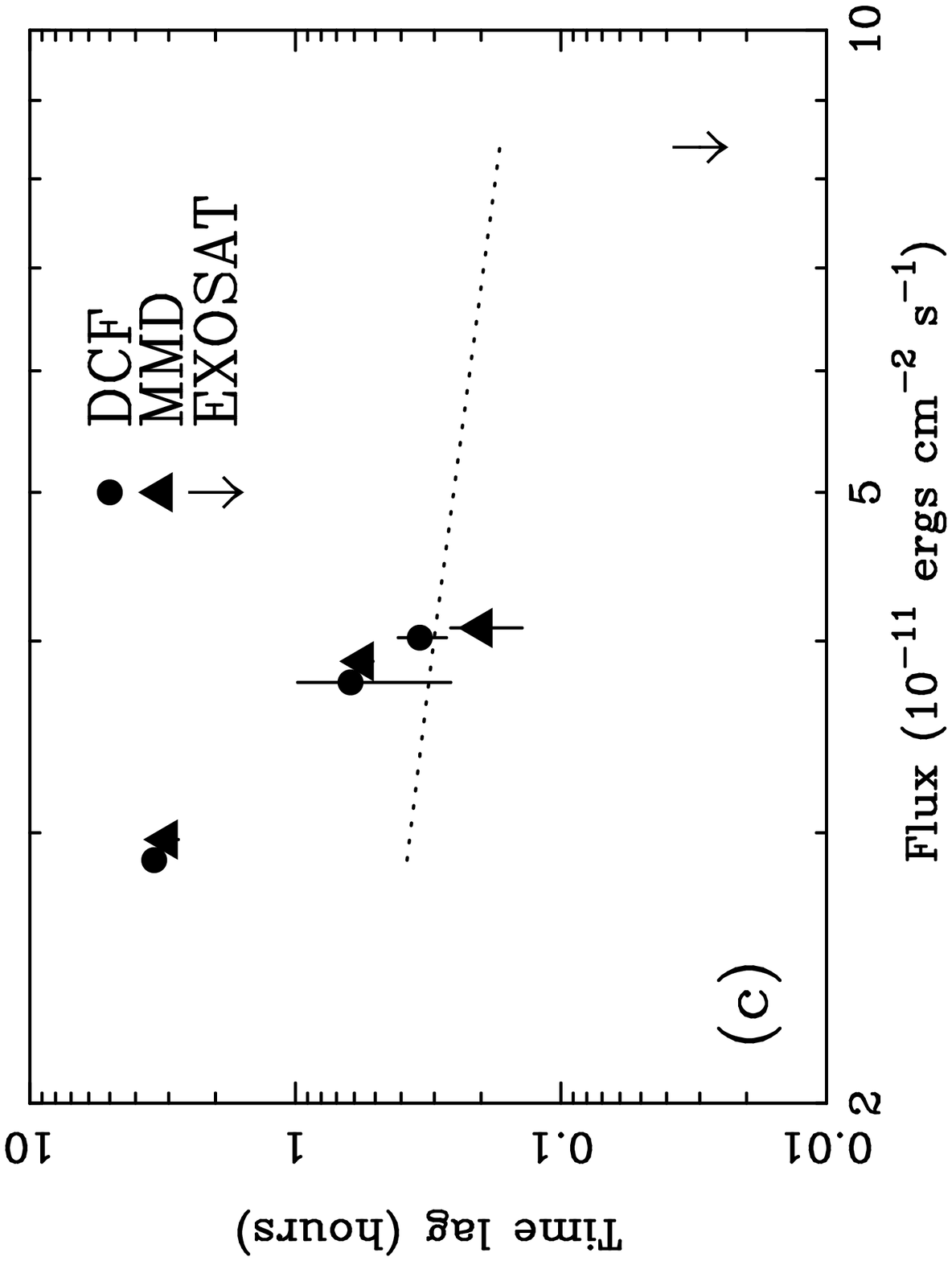}{0cm}{270}{40}{40}{-140}{40}
\end{figure}

\clearpage
\begin{figure}
\plotfiddle{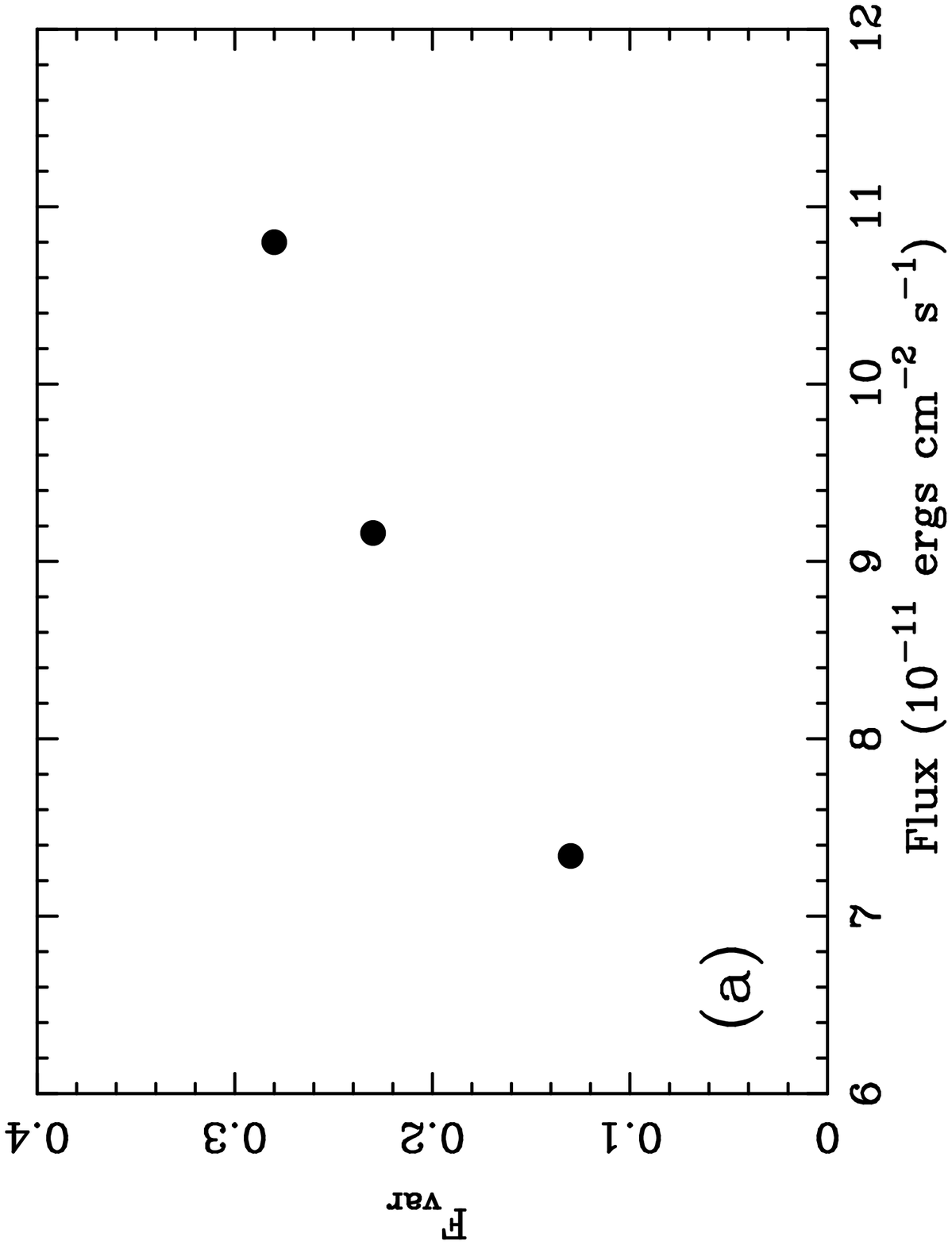}{0cm}{270}{40}{40}{-140}{320}
\plotfiddle{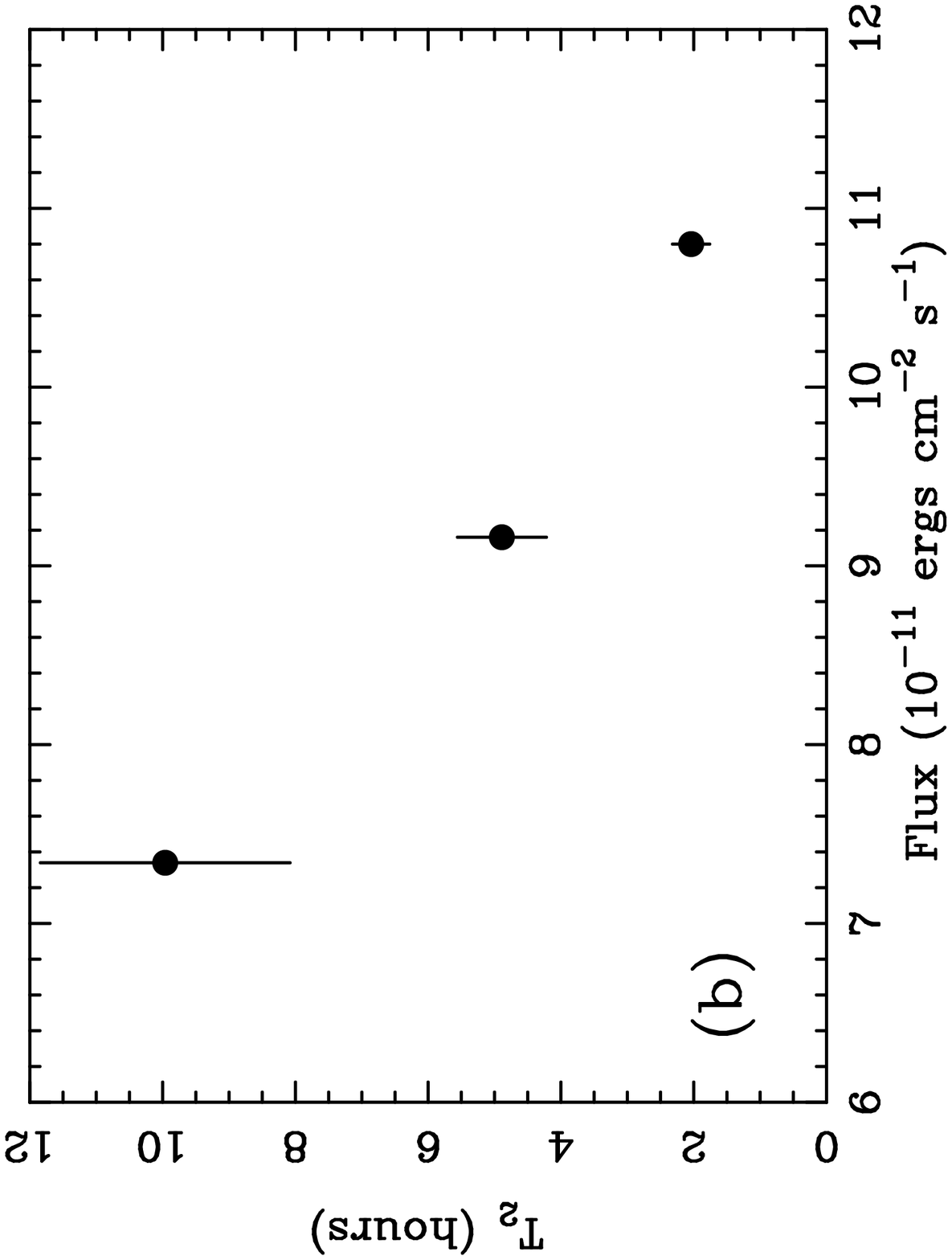}{0cm}{270}{40}{40}{-140}{180}
\end{figure}


\clearpage

\begin{thebibliography}{}

\bibitem[Boella et al. 1997a]{1997A&AS..122..299B} Boella, G., Butler,
R. C., Perola, G. C., Piro, L., Scarsi, L., \& Bleeker,
J. A. M. 1997a, A\&AS, 122, 299

\bibitem[Boella et al. 1997b]{1997A&AS..122..327B} Boella, G., et
al. 1997b, A\&AS, 122, 327

\bibitem[Brinkmann & Siebert]{1999bllp.confE.15B} Brinkmann, W., \& 
Siebert, J., 1999, Proc. of the BL Lac Phenomenon meeting, Turku, 
Finland, 22-26 June 1998, ed. L.O. Takalo, PASP Conf. Series, 159, 168 

\bibitem[Chadwick P.M. et al. 1998]{1999ApJ...513..161C} Chadwick, P.M., et 
al. 1999, \apj, 513, 161 

\bibitem[Chiaberge & Ghisellini 1999]{1999MNRAS...306..551C} Chiaberge, 
M., \& Ghisellini, G., 1999, \mnras, 306, 551

\bibitem[Chiappetti L. et al. 1999]{dummy} Chiappetti L., et al. 1999, 
\apj, 521, in press

\bibitem[Edelson 1992]{1992ApJ...401..516E} Edelson, R. A. 1992, \apj,
401, 516

\bibitem[Edelson et al. 1999]{1999ApJ...514..682} Edelson, R., \& Nandra, K. 
1999, 514, 682

\bibitem[Edelson et al. 1995]{1995ApJ...438..120E} Edelson, R. A., et
al. 1995, \apj , 438, 120

\bibitem[Edelson & Krolik 1988]{1988ApJ...333..646E} Edelson, R. A., \& 
Krolik, J. H. 1988, \apj, 333, 646

\bibitem[Frontera et al. 1997]{1997A&AS..122..357F} Frontera, F.,
Costa,E., Dal Fiume, D., Feroci, M., Nicastro, L., Orlandini, M.,
Palazzi, E., \& Zavattini, G. 1997, A\&AS, 122, 357

\bibitem[Georganpoulos & Marscher 1998]{1998ApJL...506..11G}
Georganopoulos, M. \& Marscher, A. P. 1998, \apj, 506, L11

\bibitem[Giommi et al. 1998]{1998A&A...333L...5G} Giommi, P., et
al. 1998, \aap, 333, L5

\bibitem[Hayashida et al. 1998]{1998ApJ...500..642} Hayashida, K. et al.
1998, \apj, 500, 642

\bibitem[Hufnagel & Bregman 1992]{1992ApJ...386..473H} Hufnagel, B. R., \& 
Bregman, J. N. 1992, \apj, 386, 473



\bibitem[Makino et al. 1996]{1996rftu.proc..413M} Makino, F., et al. 
1996, R\"ongtenstrahlung from The Universe, Ed.  Zimmermann, H.U., Trumper, 
J.E., Yorke, H., MPE Report 263, 413

\bibitem[Manzo et al. 1997]{1997A&AS..122..341M} Manzo, G., Giarrusso,
S., Santangelo, A., Ciralli, F., Fazio, G., Piraino, S., \& Segreto,
A. 1997, A\&AS, 122, 341

\bibitem[Maoz & Netzer 1989]{1989MNRAS...236..21M} Maoz, D., \& Netzer, 
H. 1989, MNRAS, 236, 21

\bibitem[Marscher & Gear 1985]{1985ApJ...298..114} Marscher, A. P., \&
Gear, W. K. 1985, \apj, 298, 114

\bibitem[Mastichiadis & Kirk 1997]{1997A&A...320..19M} Mastichiadis, A.,
  \& Kirk, J. M. 1997, \aap, 320, 19

\bibitem[McHardy 1999] {1998bllp.confE.102M} McHardy, I. M. 1999, 
Proc. of the BL Lac Phenomenon meeting, Turku, Finland, 22-26 June 1998,
ed. L.O. Takalo, PASP Conf. Series, 159, 155

\bibitem[Padovani & Giommi 1995]{1995ApJ...444..567P} Padovani, P., \&
Giommi, P. 1995, \apj , 444, 567

\bibitem[Paltani 1999]{dummy} Paltani, S. 1999, 
Proc. of the BL Lac Phenomenon meeting, Turku, Finland, 22-26 June 1998,
ed. L.O. Takalo, PASP Conf. Series, 159, 293

\bibitem[Parmar et al. 1997]{1997A&AS...122..309P} Parmar, A. N., et
al. 1997, A\&AS, 122, 309

\bibitem[Pesce et al. 1997]{1997ApJ...486..770P} Pesce, J. E., et al. 
1997, \apj, 486, 770

\bibitem[Peterson et al. 1998]{1998PASP...110..660P} Peterson B. M., 
Wanders I., Horne K., Collier S., Alexander T., \& Maoz D. 1998, \pasp, 
110, 660

\bibitem[Pian et al. 1997]{1997ApJ...486..784P} Pian, E., et al. 1997, 
\apj, 486, 784

\bibitem[Press 1978]{dummy} Press, W. 1978, Comments on Astrophysics, 7,
103

\bibitem[Press et al. 1992]{dummy} Press, W. et al. 1992, Numerical
Recipes: The Art of Scientific Computing, 2nd Edition, (Cambridge  
University Press)

\bibitem[Rodr\'{\i}guez-Pascual, P.M., et al. 1997]{1997ApJS...110..9R} 
Rodr\'{\i}guez-Pascual, P.M., et al. 1997, ApJS, 110, 9

\bibitem[Sreekumar & Vestrand 1997]{1997IAUC.6774....2S} Sreekumar,
P., \& Vestrand, W. T. 1997, \iaucirc , 6774, 2

\bibitem[Stella \& Angelini 1993]{dummy} Stella, L., \& Angelini, L. 
1993, XRONOS, A Timing Analysis Software Package, User's Guide, Version 3.01 

\bibitem[Tagliaferri et al. 1996]{1996ApJ...465..181T} Tagliaferri, G.,   
Bao, G., Israel, L., Stella, L., \& Treves, A., 1996, \apj, 465, 181

\bibitem[Tagliaferri et al. 1991]{1991ApJ...380..78T} Tagliaferri, G., 
Stella, L., Maraschi, L., Treves, A., \& Celotti, A., 1991, \apj, 380, 78

\bibitem[Takahashi et al. 1996]{1996ApJ...470L..89T} Takahashi, T., et 
al. 1996,  \apjl, 470, L89 

\bibitem[Tanaka et al. 1994] {1994PASJL...46..37T} Tanaka, Y., Inoue, H.,
   \& Holt, S.S. 1994, PASJ, 46, L37

\bibitem[Tavecchio et al. 1998]{1998ApJ...509..608T} Tavecchio, F., 
Maraschi, L., \& Ghisellini, G., 1998, ApJ, 509, 608

\bibitem[Treves 1999]{1998bllp.confE.128T} Treves, A., et al., 1999, 
Proc. of the BL Lac Phenomenon meeting, Turku, Finland, 22-26 June 1998, 
Ed. L.O. Takalo, PASP Conf. Series, 159, 184

\bibitem[Urry et al. 1997]{1997ApJ...486..799U} Urry, C. M., et
al. 1997, \apj , 486, 799

\bibitem[Vestrand, Stacy, & Sreekumar 1995]{1995ApJ...454L..93V}
Vestrand, W. T., Stacy, J. G., \& Sreekumar, P. 1995, \apjl , 454, L93

\end{thebibliography}
\end{document}